\documentclass[sigconf, nonacm]{acmart}
\usepackage{subcaption}
\usepackage[ruled,linesnumbered,noresetcount]{algorithm2e}
\usepackage{tikz}
\usepackage{multirow}
\usepackage{makecell}
\usepackage{adjustbox}
\usepackage{enumitem}

\makeatletter
\patchcmd{\@algocf@start}
  {-1.5em}
  {0pt}
  {}{}
\makeatother

\newcommand\vldbdoi{XX.XX/XXX.XX}

\newcommand\vldbavailabilityurl{URL_TO_YOUR_ARTIFACTS}

\let\oldnl\nl
\newcommand{\nonl}{\renewcommand{\nl}{\let\nl\oldnl}}

\newcommand{\buf}{\mathsf{buf}}
\newcommand{\tx}{\mathsf{tx}}
\newcommand{\acs}{\texttt{ACS}}
\newcommand{\aaba}{\texttt{AABA}}
\newcommand{\gbc}{\texttt{GBC}}
\newcommand{\acsq}{\Pi_{acsq}}

\newcommand{\rev}[1]{{\textcolor{black}{#1}}}
\newenvironment{revpara}{\par\color{black}}{\par}

\def\proto{Falcon}

\begin{document}

\title{\proto: Advancing Asynchronous BFT Consensus for Lower Latency and Enhanced Throughput}

\author{Xiaohai Dai}
\affiliation{
  \institution{HUST$^{\dagger}$}
}
\email{xhdai@hust.edu.cn}

\author{Chaozheng Ding}
\affiliation{
  \institution{HUST$^{\dagger}$}
}
\email{chaozhengding@hust.edu.cn}

\author{Wei Li$^{\S}$}
\affiliation{
  \institution{USYD$^{\ddagger}$}
}
\email{weiwilson.li@sydney.edu.au}

\author{Jiang Xiao}
\affiliation{
  \institution{HUST$^{\dagger}$}
}
\email{jiangxiao@hust.edu.cn}

\author{Bolin Zhang}
\affiliation{
  \institution{CMU$^*$}
}
\email{bolinz@andrew.cmu.edu}

\author{Chen Yu}
\affiliation{
  \institution{HUST$^{\dagger}$}
}
\email{yuchen@hust.edu.cn}

\author{Albert Y. Zomaya}
\affiliation{
  \institution{USYD$^{\ddagger}$}
}
\email{albert.zomaya@sydney.edu.au}

\author{Hai Jin}
\affiliation{
  \institution{HUST$^{\dagger}$}
}
\email{hjin@hust.edu.cn}

\settopmatter{authorsperrow=4}

\begin{abstract}
Asynchronous \textit{Byzantine Fault Tolerant} (BFT) consensus protocols have garnered significant attention with the rise of blockchain technology.
A typical asynchronous protocol is designed by executing sequential instances of the \textit{Asynchronous Common Sub-seQuence} (ACSQ). The ACSQ protocol consists of two primary components: the \textit{Asynchronous Common Subset} (ACS) protocol and a block sorting mechanism, with the ACS protocol comprising two stages: broadcast and agreement.
However, current protocols encounter three critical issues: high latency arising from the execution of the agreement stage, \rev{latency instability due to the integral-sorting mechanism}, and reduced throughput caused by block discarding. 

To address these issues, we propose \proto, an asynchronous BFT protocol that achieves low latency and enhanced throughput. 
\proto~introduces a novel broadcast protocol, \textit{Graded Broadcast} (GBC), which enables a block to be included in the \acs~set directly, bypassing the agreement stage and thereby reducing latency.
To ensure safety, \proto~incorporates a new binary agreement protocol called \textit{Asymmetrical Asynchronous Binary Agreement} (AABA), designed to complement GBC.
\rev{Additionally, \proto~employs a partial-sorting mechanism, allowing continuous rather than simultaneous block sorting, enhancing latency stability.}
\rev{Finally, we incorporate an agreement trigger that, before its activation, enables nodes to wait for more blocks to be delivered and committed, thereby enhancing throughput.}
We conduct a series of experiments to evaluate \proto, demonstrating its superior performance.
\end{abstract}

\maketitle


\renewcommand\thefootnote{}

\footnote{
\noindent $^{\S}$Corresponding author. \\
\noindent $^{\dagger}$National Engineering Research Center for Big Data Technology and System, Services Computing Technology and System Lab, Cluster and Grid Computing Lab, School of Computer Science and Technology, Huazhong University of Science and Technology. \\
\noindent $^{\ddagger}$School of Computer Science, The University of Sydney. \\
\noindent $^*$Language Technologies Institute, Carnegie Mellon University. \\
\noindent This work is licensed under the Creative Commons BY-NC-ND 4.0 International License. Visit \url{https://creativecommons.org/licenses/by-nc-nd/4.0/} to view a copy of this license. For any use beyond those covered by this license, obtain permission by emailing \href{mailto:info@vldb.org}{info@vldb.org}. Copyright is held by the owner/author(s). Publication rights licensed to the VLDB Endowment.
}

\ifdefempty{\vldbavailabilityurl}{}{
\vspace{-0.3cm}
\begingroup\small\noindent\raggedright\textbf{PVLDB Artifact Availability:}\\
The source code, data, and/or other artifacts have been made available at \url{https://github.com/CGCL-codes/falcon}.
\endgroup
}


\section{Introduction}
With the rising prominence of blockchain technology~\cite{wu2022flexchain, peng2022neuchain, amiri2022qanaat}, \textit{Byzantine Fault Tolerant} (BFT) consensus has garnered significant attention from both academia and industry~\cite{xiao2020survey, wang2022bft}.
The BFT consensus protocols ensure agreement on data and facilitate implementing the \textit{State Machine Replication} (SMR) among distributed nodes, some of which may be malicious, known as Byzantine nodes.
These protocols are categorized by timing assumptions into synchronous protocols, partially synchronous, and asynchronous types.
Given that synchronous and partially-synchronous protocols are susceptible to network attacks~\cite{miller2016honey, guo2020dumbo}, recent research has increasingly focused on developing asynchronous protocols~\cite{duan2018beat, zhang2022pace}.

\subsection{Asynchronous BFT}\label{sec:cons_async_proto}
An asynchronous BFT protocol can be constructed by executing consecutive instances of \textit{Asynchronous Common Sub-seQuence} (ACSQ), each generating a block sequence that is written to an append-only vector for execution.
An ACSQ protocol comprises two components: an \textit{Asynchronous Common Subset} (ACS) protocol and a block sorting mechanism.
The ACS protocol produces a consistent set of blocks, referred to as the \acs~set, which are then sorted into a sequence by the block sorting mechanism.
The sorting of blocks is essentially the process of writing these blocks to the append-only vector, also known as committing blocks.
\rev{A detailed explanation of these operation terms is provided in Table~\ref{tb:notation} and Section~\ref{sec:termino}.}

The construction of ACS typically unfolds in two stages: the broadcast stage and the agreement stage.
During the broadcast stage, each node disseminates blocks using the \textit{Reliable Broadcast} (RBC) protocol~\cite{bracha1987asynchronous, cachin2005asynchronous}.
The agreement stage diverges into two distinct paradigms: the BKR paradigm~\cite{ben1994asynchronous} and the CKPS paradigm~\cite{cachin2001secure}.

\begin{figure}[!tp]
    \centering
    \begin{subfigure}{\linewidth}
    \centering
        \includegraphics[width=0.97\linewidth]{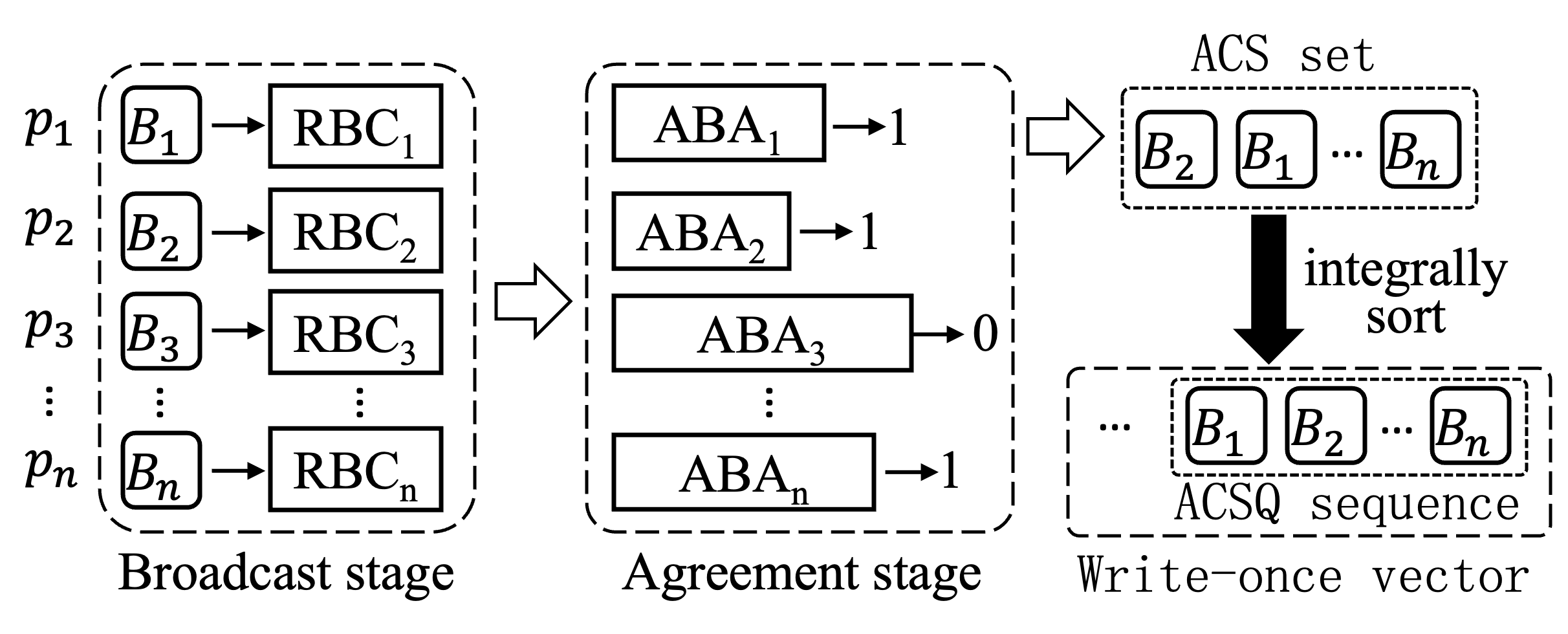}
        \vspace{-0.2cm}
        \caption{An ACSQ instance utilizing the BKR paradigm}
        \label{fig:struc-bkr}
    \end{subfigure}
    \begin{subfigure}{\linewidth}
    \centering
        \includegraphics[width=\linewidth]{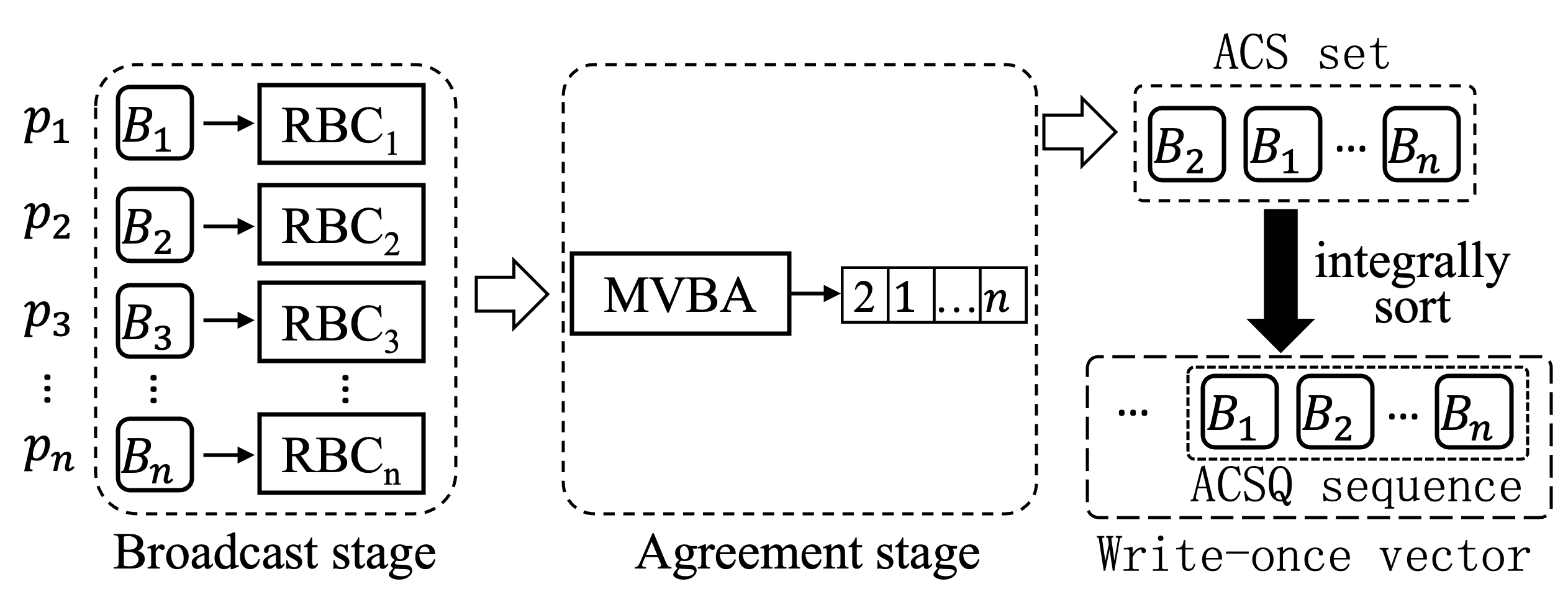}
        \vspace{-0.6cm}
        \caption{An ACSQ instance utilizing the CKPS paradigm}
        \vspace{-0.4cm}
        \label{fig:struc-ckps}
    \end{subfigure}
    \caption{The structure of ACSQ protocols}
    \label{fig:struc-bft}
\end{figure}

An ACSQ protocol utilizing the BKR paradigm is illustrated in Figure~\ref{fig:struc-bkr}, where an \textit{Asynchronous Binary Agreement} (ABA) instance is executed for each block.
\rev{ABA is used to reach agreement on binary values. Each node inputs either $0$ or $1$ into ABA, which then produces a consistent output of $0$ or $1$. In the BKR paradigm, an output of $1$ indicates that the corresponding block should be included in the \acs~set, while an output of $0$ signals rejection.}
As shown in Figure~\ref{fig:struc-bkr}, blocks $B_2$, $B_1$, ..., and $B_n$ are included in the \acs~set, whereas $B_3$ is excluded. 
On the other hand, the CKPS paradigm, depicted in Figure~\ref{fig:struc-ckps}, employs a single \textit{Multi-valued Validated Byzantine Agreement} (MVBA) instance during the agreement stage.
The MVBA protocol takes an array of block numbers (i.e., the identifiers of their creators) as input\footnote{More accurately, the input is an array of block numbers plus proofs~\cite{guo2020dumbo}.} and produces an output array that determines which blocks should be included in the \acs~set.
Following the agreement stage, blocks within the \acs~set are sorted by block numbers in both the BKR and CKPS paradigms.

However, existing protocols encounter three primary issues:

\begin{figure}[!tp]
    \centering
     \begin{subfigure}{0.48\linewidth}
    \centering
        \includegraphics[width=0.7\linewidth]{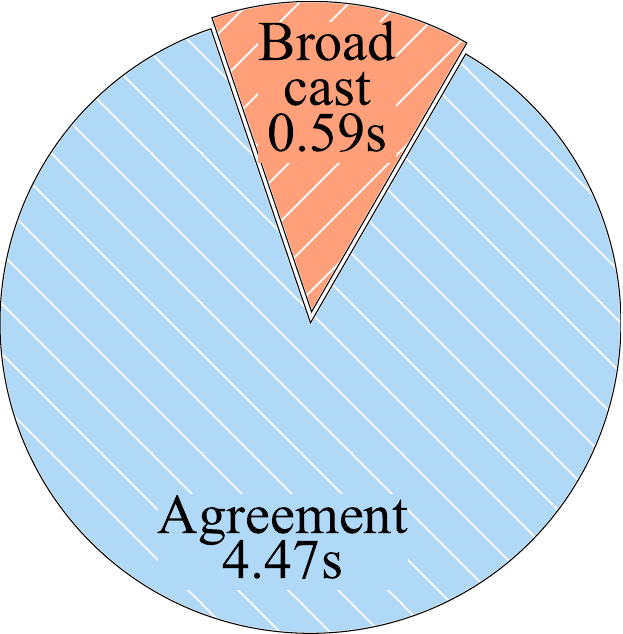}
        \vspace{-0.1cm}
        \caption{\rev{BKR-ACS}}
        \vspace{-0.4cm}
        \label{fig:pie-hbbft}
    \end{subfigure}
    \begin{subfigure}{0.48\linewidth}
    \centering
        \includegraphics[width=0.7\linewidth]{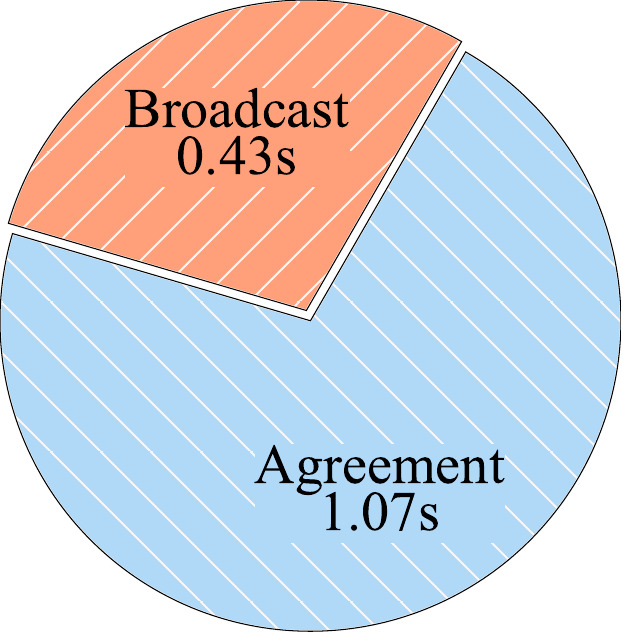}
        \vspace{-0.1cm}
        \caption{\rev{CKPS-ACS}}
        \vspace{-0.4cm}
        \label{fig:pie-sdumbo}
    \end{subfigure}
    \caption{\rev{Latency decomposition}}
    \vspace{-0.1cm}
    \label{fig:lat-decomp-pre}
\end{figure}

\vspace{-0.15cm}
\subsubsection{Issue 1: high latency}
As analyzed in \cite{guo2020dumbo}, the latency of the BKR-ACS protocol (i.e., the ACS protocol utilizing the BKR paradigm) is defined by the sum of two stages: $t_{r}+\log(n) \cdot t_{a}$, where $t_{r}$ represents the latency for a RBC instance, $t_{a}$ for an ABA instance, and $n$ denotes the node count. BKR-ACS's latency can be quite high for large $n$.
As for the CKPS-ACS protocol, it has a latency of $t_{r}+ t_{m}$, where $t_{m}$ denotes the latency of an MVBA instance.
While this protocol eliminates the logarithmic term, $t_{m}$ introduces a significant constant term. Specifically, even the state-of-the-art MVBA protocol, sMVBA~\cite{guo2022speeding}, has a best-case latency of six communication rounds, compared to just three rounds for $t_{r}$.
We conduct preliminary experiments with 16 nodes to analyze the latency at different stages of the ACS protocol, with HBBFT~\cite{miller2016honey} and Dumbo~\cite{guo2020dumbo} representing BKR-ACS and CKPS-ACS. The detailed experimental setup is provided in Section~\ref{sec:impl-set}, and results are presented in Figure~\ref{fig:lat-decomp-pre}. As shown, in both paradigms, \textbf{the agreement stage contributes significantly to the overall latency}.

\begin{figure}
    \centering
        \includegraphics[width=0.7\linewidth]{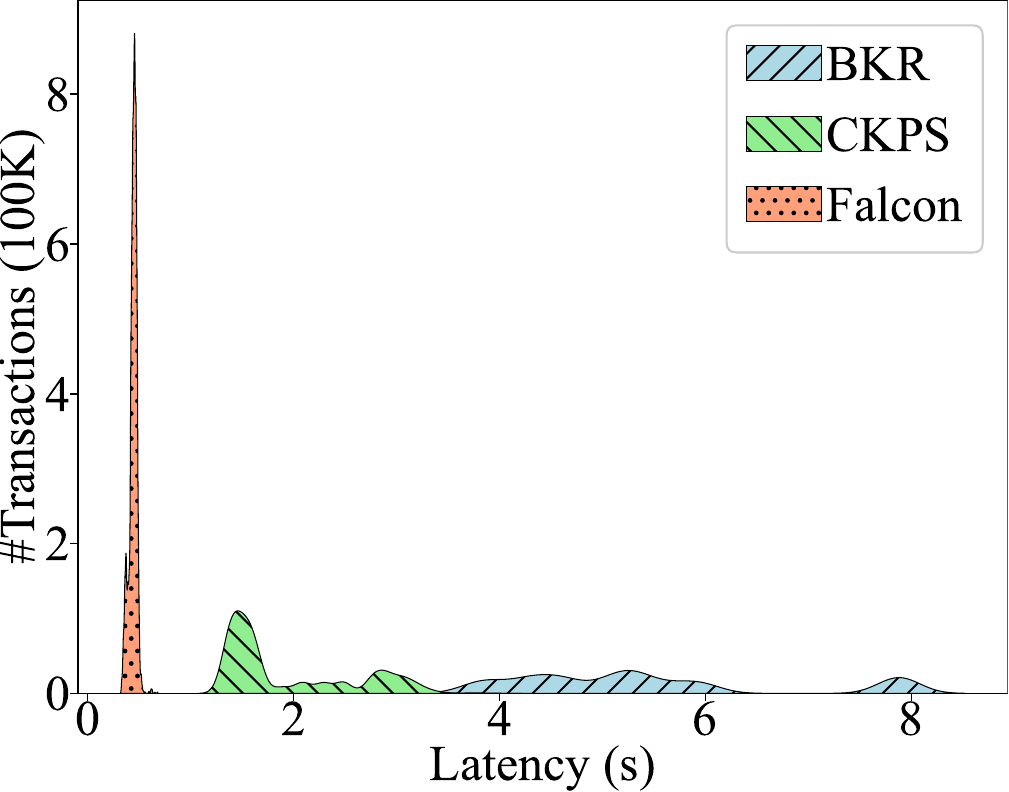}
        \vspace{-0.4cm}
        \caption{\rev{Latency distribution}}
        \vspace{-0.4cm}
        \label{fig:lat-dist}
\end{figure}

\vspace{-0.15cm}
\subsubsection{\rev{Issue 2: latency instability}}
As illustrated in Figure~\ref{fig:struc-bft}, blocks in the \acs~set are committed as a whole only after all blocks have been decided from the agreement stage.
It is crucial to distinguish between the terms ``decide'' and ``commit'': deciding blocks refers to determining their inclusion or exclusion from the \acs~set, while committing blocks involves sorting those that are included.
In the protocol utilizing BKR, blocks in the \acs~set cannot be committed until the last ABA instance outputs.
In the protocol using CKPS, blocks are simultaneously included or excluded from the \acs~set after the MVBA instance outputs, and all blocks in this set are then sorted together.
\rev{We refer to the mechanism of performing sorting only after all blocks have been decided as ``integral sorting.''
\textbf{This integral-sorting mechanism introduces significant latency instability}, as demonstrated by the experiments on transaction latency distribution shown in Figure~\ref{fig:lat-dist}.
Specifically, the latency of BKR is widely distributed within the range of [3.1s, 8.3s], while the latency of CKPS ranges from 1.3s to 3.9s. Such latency instability can result in poor user experience in higher-layer applications.}

\vspace{-0.15cm}
\subsubsection{Issue 3: reduced throughput}
In existing ACSQ constructions, an ACSQ instance outputs only a portion of all the broadcast blocks, typically $n-f$ (where $f$ denotes fault tolerance), even when all nodes are correct and all blocks are well-broadcast.
Specifically, in the BKR paradigm, once $n-f$ ABA instances output $1$, nodes input `0' into the remaining $f$ ABA instances, causing them to output `0' and resulting in the corresponding blocks being excluded from the \acs~set and discarded.
In the CKPS paradigm, each input array for the MVBA instance contains only $n-f$ block numbers, leading to an output array that also comprises $n-f$ block numbers. Consequently, a maximum of $n-f$ blocks can be committed.
\textbf{This leads to reduced throughput, as up to $f$ blocks may be discarded}.

\vspace{-0.15cm}
\subsection{Our solution \& evaluation}\label{sec:our-sol}
To address the above issues, we propose \proto, which delivers both low and stable latency as well as enhanced throughput.
At a high level, \proto~utilizes ABA instances for the agreement stage, and is designed based on our three insights:
\vspace{-0.03cm}
\begin{itemize}[left=6pt]
    \item \textbf{Insight 1:} If a node $p_i$ delivering a block $B_k$ can lead the corresponding ABA instance to eventually output $1$, then $p_i$ can include $B_k$ in its \acs~set as soon as $B_k$ is delivered. 
    \item \textbf{Insight 2:} For a specific block $B_k$, if a node $p_i$ has decided on all blocks with smaller numbers, $p_i$ can commit $B_k$ immediately when it is included in the \acs~set. 
    \item \textbf{Insight 3:} \rev{A longer wait during the broadcast stage helps prevent premature abandonment of blocks, thereby contributing to the committing of more blocks.}
\end{itemize}

\vspace{-0.15cm}
\subsubsection{Solution to issue 1.}
Building on \textbf{insight 1}, we propose a new broadcast protocol, \textit{Graded Broadcast} (GBC), for the broadcast stage, along with a novel binary agreement protocol, \textit{Asymmetrical ABA} (AABA), for the agreement stage.
With GBC, a node can deliver blocks with two grades, $1$ and $2$.
\rev{The intuition behind introducing grades is that we want a correct node to deliver a block with grade $2$ only after $f+1$ correct nodes have already delivered it with a lower grade (i.e., grade $1$). These nodes that deliver with grade $1$ will input $1$ in the subsequent AABA instance. In this case, AABA's biased-validity property further ensures that the AABA instance will output $1$, thus satisfying the condition in \textbf{insight 1}.}

In \proto, once a node delivers a block with grade $2$, it can immediately include that block into its \acs~set.
For nodes that deliver the block with grade $1$, they will input $1$ to the AABA instance.
Additionally, if a node outputs $1$ from the AABA instance, it will also include the corresponding block in the \acs~set.
Furthermore, we incorporate a shortcut mechanism to the AABA protocol, allowing it to output more quickly if all nodes input $0$.

In a favorable situation, each node can deliver all blocks with grade $2$ during the broadcast stage and include them directly in its \acs~set, bypassing the time-consuming agreement stage.
This significantly reduces latency and effectively addresses \textbf{issue 1}.

\vspace{-0.15cm}
\subsubsection{Solution to issue 2.}
Based on \textbf{insight 2}, we devise a partial-sorting mechanism in \proto, enabling a block to be committed directly if it is included in the \acs~set and all preceding blocks have been decided.
For example, consider three blocks, $B_1$, $B_2$, and $B_3$, created by replicas $p_1$, $p_2$, and $p_3$, within the same ACSQ instance.
If $B_1$ is decided to be included and $B_2$ is excluded, $B_3$ can be committed immediately once it is included in the \acs~set.
\rev{This eliminates the need to wait for decisions on all blocks, ensuring more stable latency and solving \textbf{issue~2}.
The preliminary experimental results presented in Figure~\ref{fig:lat-dist} indicate that \proto's transaction latency fluctuates within a narrow range of [0.3s, 0.7s].}
Furthermore, this mechanism allows blocks to be committed earlier, further reducing overall latency.

\vspace{-0.15cm}
\subsubsection{Solution to issue 3.}
Following \textbf{insight 3}, \rev{we introduce an \textit{agreement trigger} that allows nodes to wait for more blocks to be delivered.} 
Specifically, once a node has delivered $n-f$ blocks with grade $2$ from the GBC instances, \rev{it checks whether the trigger has been activated}.
If so, the node proceeds directly to the agreement stage; otherwise, it waits until either all blocks are delivered with grade $2$ or \rev{the trigger is activated}.
If all blocks are delivered with grade $2$ before \rev{the trigger activation}, they will be included in the \acs~set, ensuring no blocks are discarded and resolving \textbf{issue 3}.

\rev{The trigger is activated by system events, particularly the grade-$2$ delivery of a block from the next ACSQ instance. 
In favorable situations, blocks from the current ACSQ will always be delivered before those from the next ACSQ, keeping the trigger inactive. Conversely, if the blocks from the next ACSQ are delivered first, it indicates an unfavorable situation, prompting the system to transition to the agreement stage based on the trigger activation.}



\vspace{-0.15cm}
\subsubsection{Evaluation}
To evaluate \proto's performance, we implement a prototype system and \rev{conduct a series of comparisons with four representative protocols: HBBFT~\cite{miller2016honey}, Dumbo~\cite{guo2020dumbo}, MyTumbler~\cite{liu2023flexible}, and Narwhal\&Tusk~\cite{danezis2022narwhal}.} 
We examine both favorable situations, where all nodes are correct, and unfavorable ones, which include some faulty nodes.
The experimental results show that \proto~achieves lower latency compared to other protocols in both situations. In favorable situations, this advantage arises from its use of GBC for direct block committing, bypassing the agreement stage. 
In unfavorable situations, \proto's low latency is attributed to the AABA protocol's shortcut mechanism, enabling faster outputs. 

\proto~also surpasses others in throughput, as \rev{it facilitates the delivery and committing of more blocks before the trigger is activated}.
\rev{In terms of latency stability, \proto~achieves more stable latency due to the partial-sorting mechanism that enables continuous block committing.} By breaking down latency into different stages, we find that \proto~incurs minimal latency during both the agreement and sorting stages, further validating its effectiveness.

\section{Model and Preliminaries}
\subsection{System model}\label{sec:model}
We consider a system consisting of $n$ nodes, of which up to $f$ ($3f+1 \leq n$) can exhibit Byzantine behavior.
These Byzantine nodes are presumed to be controlled by an adversary capable of coordinating their actions.
The other nodes, termed correct, strictly adhere to the protocol.
Each node is uniquely identified by a distinct number, denoted as $p_i$ ($1 \leq i \leq n$).
The system operates over an asynchronous network, where no assumptions are made about network delays.
Furthermore, the adversary is assumed to have the ability to delay and reorder message delivery, though all messages are eventually delivered.
This assumption of an asynchronous network ensures that the consensus protocol developed on this network framework is robust enough to withstand various network manipulation attacks~\cite{miller2016honey}.
Each pair of nodes is connected via an authenticated network link.

\rev{A \textit{Public Key Infrastructure} (PKI) and a threshold signature scheme are established within the system.
The adversary is assumed to be computationally bounded, implying that it cannot compromise either the PKI or the threshold signature scheme.}

\vspace{-0.2cm}
\subsection{SMR and BFT protocols}\label{sec:smr}
We focus on a \textit{State Machine Replication} (SMR), where each node $p_i$ maintains a local vector denoted as $\mathcal{C}_i$.
Each slot within $\mathcal{C}_i$ is indexed by $r$ and is appended only.
Initially, $\mathcal{C}_i$ is empty, with $\mathcal{C}_i[r]=\bot$ for each $r$ (where $r \geq 1$).
Without loss of generality, we assume elements to be written in $\mathcal{C}_i$ are blocks, and each block comprises multiple transactions submitted by clients.
Blocks, particularly their contained transactions, in the append-only vector, are eligible for execution to change the machine state.
In the field of consensus research, we mainly focus on how to maintain the vector, without considering the execution operations after that.

A block is considered committed by $p_i$ when it is written to $\mathcal{C}_i$.
Besides, a block ready for committing is written to the first available empty slot. In other words, a block $B$ is written at $\mathcal{C}_i[r]$ only if, for each index $l$ prior $r$, $\mathcal{C}_i[l] \neq \bot$.
We define the length of a chain $\mathcal{C}_i$ as the maximum index of its non-empty slots, denoted by $\left| \mathcal{C}_i \right|$. 

\begin{figure*}[!tp]
    \centering
    \begin{minipage}[b]{0.35\linewidth}
       \centering
        \includegraphics[width=0.8\linewidth]{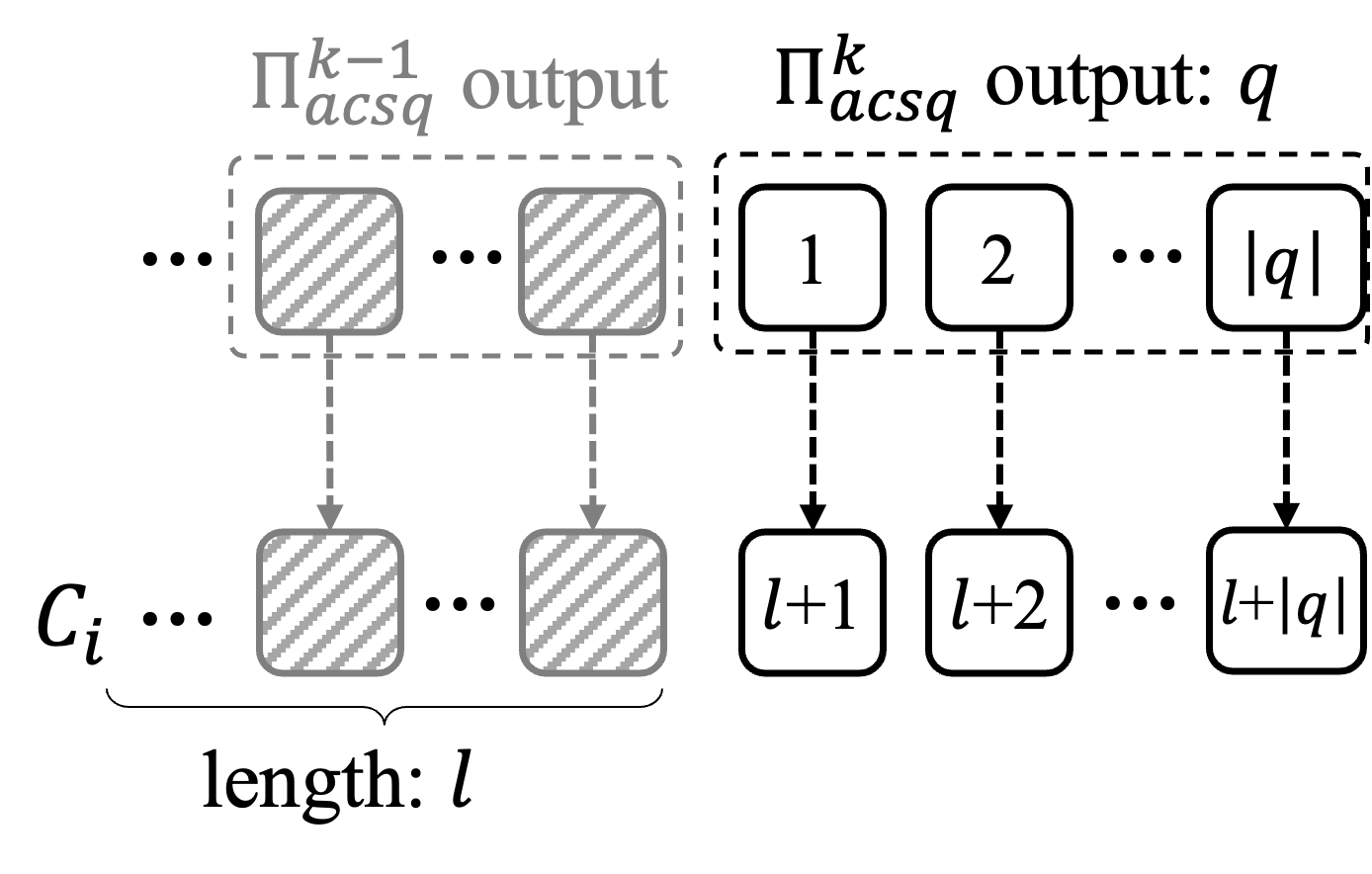}
        \vspace{-0.3cm}
        \caption{\rev{BFT construction based on ACSQ}}
        \vspace{-0.1cm}
        \label{fig:bft-construc}
    \end{minipage}
    \hspace{0.01\linewidth} 
    \begin{minipage}[b]{0.3\linewidth}
       \centering
        \includegraphics[width=0.78\linewidth]{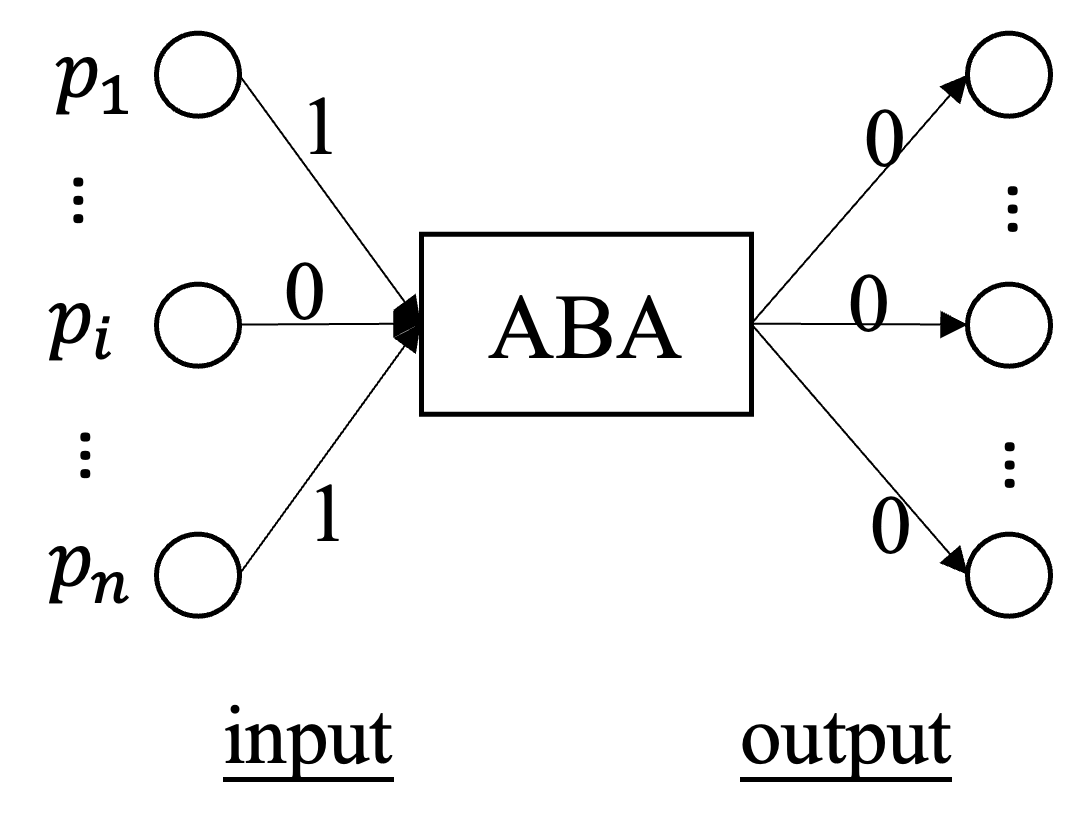}
        \vspace{-0.3cm}
        \caption{\rev{Schematic diagram of ABA}}
        \vspace{-0.1cm}
        \label{fig:aba}
    \end{minipage}
    \hspace{0.01\linewidth} 
    \begin{minipage}[b]{0.3\linewidth}
       \centering
        \includegraphics[width=\linewidth]{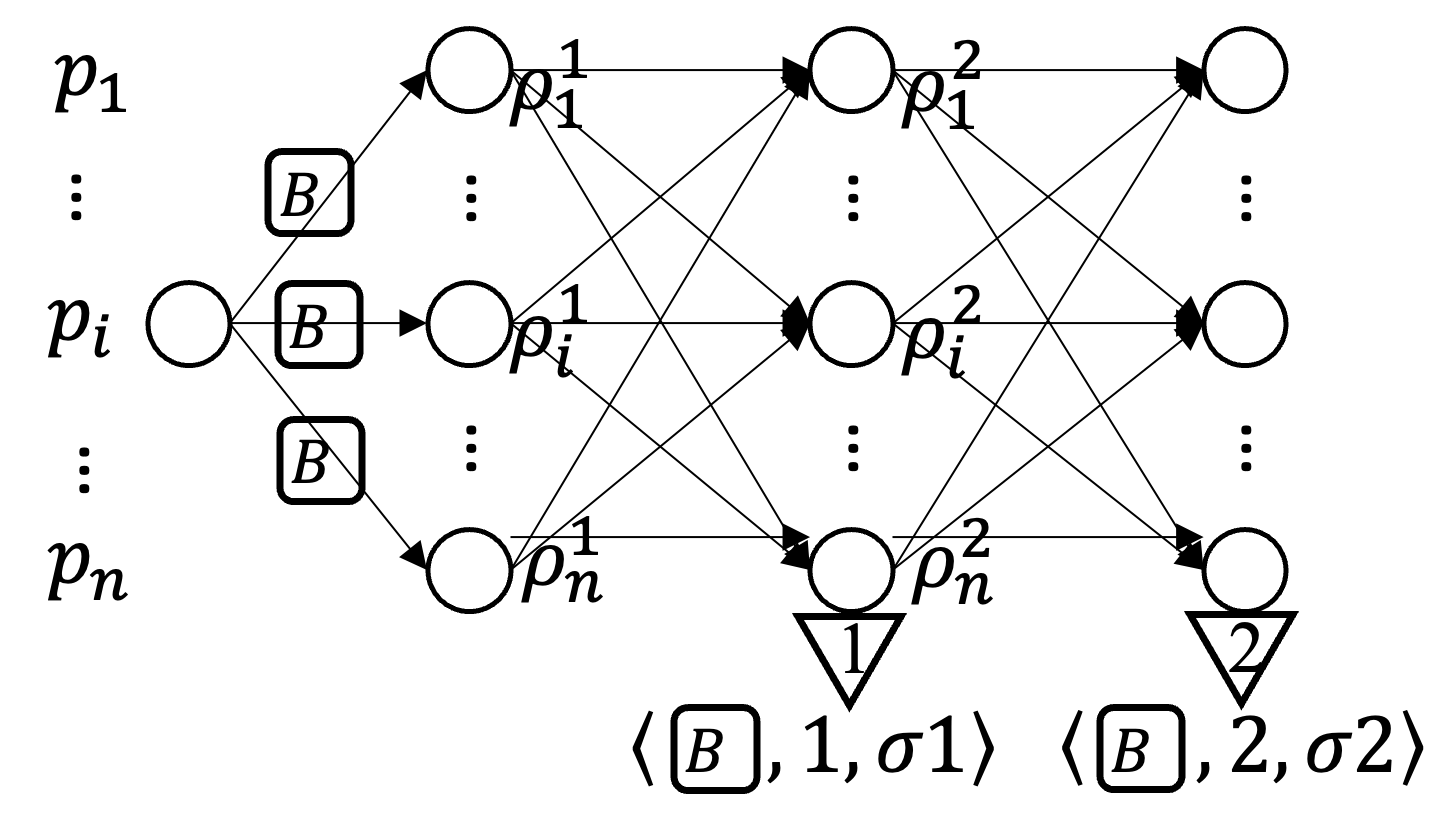}
        \vspace{-0.5cm}
        \caption{Construction of GBC}
        \vspace{-0.1cm}
        \label{fig:gbc-struc}
    \end{minipage}
\end{figure*}

Additionally, each node $p_i$ maintains a local buffer, denoted as $\buf_i$.
Transactions submitted by clients are initially cached in this buffer. 
When proposing a block, a node extracts a group of transactions, typically those at the prefix of the buffer.
Transactions that are contained in a committed block are then removed from the buffer.
Specifically, a correct BFT protocol has to satisfy the following two properties:
\begin{itemize}[left=6pt]
    \item \textbf{Safety:} For two nodes $p_i$ and $p_j$, if $\mathcal{C}_i[r] \neq \bot$ and $\mathcal{C}_j[r] \neq \bot$, then $\mathcal{C}_i[r] = \mathcal{C}_j[r]$.
    \item \textbf{Liveness:} If a transaction $\tx$ is included in every correct node's buffer, each one will eventually commit $\tx$.
\end{itemize}

\vspace{-0.4cm}
\subsection{Asynchronous common sub-sequence}\label{sec:acsq}
\subsubsection{Definition of ACSQ}
Before introducing \textit{Asynchronous Common Sub-seQuence} (ACSQ) protocol, we first introduce a basic concept named \textit{Asynchronous Common Subset} (ACS)\rev{~\cite{ben1994asynchronous, cachin2001secure}}. Generally speaking, the ACS protocol is a single-shot protocol, where each node inputs a block and then outputs an identical set of blocks.
The ACS protocol must satisfy the following three properties:
\begin{itemize}[left=6pt]
    \item\rev{\textbf{Agreement:} If a correct node outputs a set $s$, then every correct node outputs $s$.}
    \item \textbf{Validity:} If a correct node outputs a set $s$, then $\left| s \right| \geq n-f$.
    \item \rev{\textbf{Totality:} If each correct node receives an input, all correct nodes will eventually output.}
\end{itemize}

The ACSQ protocol can be considered a sorted version of the ACS protocol, in that the output from ACSQ is a sequence of blocks after sorting.
Concretely speaking, the ACSQ protocol is defined by the following properties:
\begin{itemize}[left=6pt]
    \item \rev{\textbf{Agreement:} If a correct node outputs a sequence $q$, then every correct node outputs $q$}.
    \item \textbf{Validity:} If a correct node outputs a sequence $q$, then $\left| q \right| \geq n\text{-}f$.
    \item \rev{\textbf{Totality:} If each correct node receives an input, all correct nodes will eventually output.}
\end{itemize}

The validity property of ACSQ mandates the inclusion of at least $n-f$ blocks. 
However, as discussed in \textbf{issue 3} of Section~\ref{sec:cons_async_proto}, this permits the discarding of up to $f$ blocks, even under favorable situations.
To address this, we introduce a new property termed `optimistic validity' for ACSQ, defined as follows:
\begin{itemize}[left=6pt]
    \item \textbf{Optimistic validity:} If all nodes are correct and the network conditions are favorable, then each correct node can output a sequence comprising inputs from all nodes, such that $\left| q \right| = n$.
\end{itemize}
\rev{A favorable network refers to a network where, if a correct node first broadcasts block $B1$ and then broadcasts block $B2$ after delivering $B1$, all correct nodes are guaranteed to deliver $B1$ before $B2$.}

\begin{table}[t]
    \renewcommand\arraystretch{1.2}
    \centering      
    \caption{\rev{Terminology for block operations. I, II, and III denote broadcast, agreement, and sorting stages in ACSQ.}}
    \label{tb:notation}
    \begin{tabular}{m{1.7cm} m{5cm} m{0.7cm}}                      
        \hline
        \multicolumn{1}{c}{\rev{\textbf{Operation}}} & \multicolumn{1}{c}{\rev{\textbf{Explanation}}} & \multicolumn{1}{c}{\rev{\textbf{Stage}}}\\
        \hline
        \multicolumn{1}{c}{\rev{Deliver}} & \rev{Accept a block} & \rev{I} \\
        \hline
        \multicolumn{1}{c}{\rev{Decide}} & \rev{Include/exclude a block in/from \acs~set} & \rev{I or II}\\
        \hline
        \multicolumn{1}{c}{\rev{Sort/Commit}} & \rev{Sort blocks in the \acs~set and write sorted blocks into $\mathcal{C}_i$} & \rev{III} \\
        \hline
    \end{tabular}
\end{table}

\vspace{-0.15cm}
\begin{revpara}
\subsubsection{ACSQ construction \& terminology}\label{sec:termino}
As described in Section~\ref{sec:cons_async_proto}, ACSQ can be implemented by adding a sorting stage to the ACS protocol, which itself is comprised of the broadcast and agreement stages. To clarify the terms related to block operations at each stage, we provide explanations in Table~\ref{tb:notation}. During the broadcast stage, the operation of a node accepting a block is referred to as ``deliver''. Through the agreement stage, a decision is made about a block's inclusion or exclusion in/from the \acs~set, which is termed ``decide''. Since \proto~can bypass the agreement stage, the ``decide'' operation may also be completed at the end of the broadcast stage. Blocks in the \acs~set are sorted in the sorting stage. 
The sorting process is essentially the process of writing blocks to $\mathcal{C}_i$, which is referred to as ``commit''. Therefore, we interchangeably use the terms ``sort'' and ``commit'' in this paper.
\end{revpara}


\vspace{-0.15cm}
\subsubsection{Constructing BFT from ACSQ}\label{sec:smr-acsq}
The BFT protocol can be effectively constructed by operating consecutive ACSQ instances, each denoted as $\acsq^k$.
\rev{As shown in Figure~\ref{fig:bft-construc}, a node $p_i$ writes the sequence \rev{$q$} outputted from $\acsq^k$ to $\mathcal{C}_i$ only after all outputs from previous ACSQ instances $\acsq^m$ (where $m < k$) have been written.}
The blocks in $q$ are then written to $\mathcal{C}_i$ in their original order in \rev{$q$}, specified as \rev{$\mathcal{C}_i\left[l+1 \colon l+\left| q \right|\right] = q$}, where $l$ represents the length of the chain $\mathcal{C}_i$ before writing \rev{$q$}.

Using the induction method, it is straightforward to demonstrate that the BFT construction from ACSQ upholds the safety property as outlined in Section~\ref{sec:smr}.
Concerning the liveness property, each correct node will include the transaction $\tx$ in its input block for the next $\acsq$ instance as long as $\tx$ remains uncommitted.
Consider that in instance $\acsq^k$ where $\tx$ is still uncommitted, every correct node will include $\tx$ in its block input.
Given the validity and totality properties, the output \rev{$q$} from $\acsq^k$ will contain at least $n-f$ blocks, with at least $n-2f$ of these blocks inputted by correct nodes. 
Thus, \rev{$q$} will definitely include $\tx$, ensuring $\tx$ to be committed. 
In other words, the BFT construction from ACSQ is shown to guarantee the liveness property as defined in Section~\ref{sec:smr}.

\vspace{-0.15cm}
\subsection{Asynchronous binary agreement}
The \textit{Asynchronous Binary Agreement} (ABA) protocol is recognized as one of the simplest Byzantine agreement protocols.
Within an ABA instance, each node inputs a binary value and expects a consistent binary value as output.
\rev{As exemplified in Figure~\ref{fig:aba}, $p_1$ and $p_n$ input $1$ to ABA, while $p_i$ inputs $0$. Finally, all nodes output the same bit $0$.}
To be more specific, an ABA protocol is defined by the following properties:
\begin{itemize}[left=6pt]
    \item \rev{\textbf{Agreement:} If a correct node outputs $b$, then every correct node outputs $b$.}
    \item \textbf{Validity:} If a correct node outputs $b$, then at least one correct node inputs $b$.
    \item \rev{\textbf{Termination:} If each correct node receives an input, all correct nodes will eventually output.}
\end{itemize}

Over the past few decades, various constructions of ABA protocols have been developed~\cite{ben1983another, friedman2005simple, mostefaoui2014signature, abraham2022efficient}.
We utilize the ABA protocol in a black-box manner, allowing for the easy adoption of any existing ABA protocol within our ACSQ construction.

\section{Building Blocks}



\subsection{Graded broadcast (GBC)}\label{sec:gbc}

\begin{revpara}
\subsubsection{Intuition behind GBC}\label{sec:intuit-gbc}
As outlined in Section~\ref{sec:our-sol}, \proto~allows nodes to include delivered blocks to the \acs~set at the end of the broadcast stage, thus reducing latency. 
However, due to network asynchrony, some nodes may not deliver the block by then and must rely on the binary agreement for a decision. 
To ensure that the binary agreement decides to include the block to the \acs\ set (i.e., outputs $1$), a mechanism is required. This mechanism guarantees that if a correct node delivers the block, a sufficient number of correct nodes must have delivered it \textit{in some weaker form}. This directs them to input $1$ into the binary agreement. The GBC broadcast protocol is used to implement this mechanism. It guarantees that if a correct node delivers a block with grade $2$, then at least $f+1$ correct nodes have delivered the block with a lower grade of $1$.
\end{revpara}

\vspace{-0.1cm}
\subsubsection{Definition of GBC}\label{sec:def-gbc}
Similar to other broadcast protocols, a node in GBC acts as the broadcaster to disseminate a block, while the others act to deliver the block.
The block can be delivered with two grades, namely $1$ and $2$. Besides a block, a node will also deliver proof that certifies the receipt/delivery situation among the nodes.
Therefore, a node will deliver a block in the format of $\left<B, g, \sigma\right>$, where $B$ signifies the block data, $g$ ($g \in \{1, 2\}$)
denotes the grade, and $\sigma$ represents the proof.
In the context of GBC, we differentiate the terms `receive' and `deliver.' A node is said to receive a block $B$ once it obtains $B$ from the broadcaster. By contrast, a node is said to deliver $B$ if some accompanying proof $\sigma$ is also generated.
The GBC protocol has to satisfy the following properties:
\begin{itemize}[left=6pt]
    \item \textbf{Consistency:} If two correct nodes $p_i$ and $p_j$ deliver $\left<B_i, g_i, \sigma_i\right>$ and $\left<B_j, g_j, \sigma_j\right>$ respectively, then $B_i=B_j$.
    \item \textbf{Delivery-correlation:} If a correct node delivers $\left<B, 2, \sigma\right>$, at least $f+1$ correct nodes have delivered $B$ with grade $1$.
    \item \textbf{Receipt-correlation:} If a correct node delivers $\left<B, 1, \sigma\right>$, then at least $f+1$ correct nodes have received $B$.
\end{itemize}

\rev{Note that while GBC shares some similarities with Abraham et al.'s Gradecast~\cite{abraham2022gradecast}
and Malkhi et al.'s BBCA~\cite{malkhi2024bbca} protocols, there are also key differences. First, Gradecast is designed for synchronous networks and defines a `validity' property, whereas GBC makes no assumptions about network synchrony and does not define such a property. Additionally, Gradecast's `agreement' property requires all correct nodes to output, while GBC's `delivery-correlation' property only requires $f+1$ correct nodes to output. Second, BBCA introduces a probe mechanism that returns a \texttt{NOADOPT} result, used by BBCA-Chain for new block generation. 
GBC does not define a similar probe mechanism or \texttt{NOADOPT} result.
In addition to defining the `delivery-correlation' property, which is similar to the `complete-adopt' property in BBCA, GBC also introduces the `receipt-correlation' property, which plays a crucial role in the proof of Lemma 1 in Section~\ref{sec:rig-anal}.
}

\vspace{-0.1cm}
\subsubsection{Construction of GBC}
The construction of GBC is inspired by the normal-case protocol of PBFT~\cite{castro1999practical}.
As illustrated in Figure~\ref{fig:gbc-struc}, after receiving the block $B$, each node $p_j$ will broadcast a partial threshold signature $\rho_j^1$ on the concatenation of $B$'s digest and a tag number $1$.
After receiving $n-f$ messages of $\rho_j^1$, a node can construct a complete threshold signature $\sigma 1$ and deliver $\left<B, 1, \sigma 1\right>$.
Besides, it will further broadcast a partial threshold signature $\rho_j^2$ on the concatenation of $B$'s digest and a tag number $2$.
Similarly, a node can construct a complete threshold signature $\sigma 2$ based on $n-f$ received $\rho_j^2$ and deliver $\left<B, 2, \sigma 2\right>$.
\rev{This construction achieves all GBC properties, even in the presence of a Byzantine broadcaster.
Specifically, consider the Byzantine broadcaster sending two contradictory blocks, $B$ and $B'$. Since the delivery requires the collection of partial threshold signatures from $n-f$ nodes for the same block, and given that $n \geq 3f+1$, at most one of $B$ and $B'$ can be delivered, thus ensuring consistency.
However, a Byzantine node can remain silent, meaning it refuses to broadcast any block. Nevertheless, this only reduces the consensus throughput without compromising consistency or safety.
}



\vspace{-0.2cm}
\subsection{Asymmetrical ABA (AABA)}\label{sec:aaba}
\subsubsection{Intuition behind AABA}
Our intuition behind AABA mainly unfolds two aspects. 
First, as briefly described in Section~\ref{sec:gbc}, the GBC protocol guarantees that if a correct node delivers a block with grade 2, then at least $f+1$ correct nodes have delivered the block with grade 1. We need to design an ABA variant that ensures, under these conditions, it will ultimately decide to include the block in the \acs\ set.
Second, we hope to introduce a shortcut outputting mechanism for the situation where all nodes input $0$.
This mechanism can significantly accelerate the agreement progress when some nodes crash before the start of the broadcast stage.

\vspace{-0.1cm}
\subsubsection{Definition of AABA}\label{sec:def-aaba}
Different from ABA, AABA accepts asymmetrical inputs. 
Specifically, it accepts $0$ directly while accepting $1$ only if an externally defined value $v$ and the certified proof $\sigma$ are provided in the format of a triplet $\left<1, v, \sigma \right>$.
Also, an external validation predicate $Q$ is defined regarding $v$ and $\sigma$. In the context of \proto, $v$ is the digest of a block, $\sigma$ is the proof $\sigma1$ when delivering the block with grade $1$, while $Q$ is a threshold signature verification function.
Similar to ABA, AABA also produces the output of a bit.

AABA defines asymmetrical properties on bits of $0$ and $1$.
It requires AABA to output $1$ if at least $f+1$ correct nodes receive the input $\left<1, v, \sigma \right>$.
Furthermore, if AABA outputs $1$, at least one node, whether Byzantine or correct, must have received $\left<1, v, \sigma \right>$ such that $Q(v, \sigma)$=\texttt{true}.
Besides, AABA enables a shortcut to commit $0$: if all nodes receive the input $0$, AABA can output $0$ quickly.

To be more specific,  AABA is defined by the following properties:
\begin{itemize}[left=6pt]
    \item \rev{\textbf{Agreement:} If a correct node outputs $b$, then every correct node outputs $b$.}
    \item \textbf{$1$-validity:} If a correct node outputs $1$, then at least one node, whether Byzantine or correct, must have received the input $\left<1, v, \sigma \right>$ s.t. $Q(v, \sigma)$=\texttt{true}. 
    \item \textbf{Biased-validity:} If $f+1$ or more correct nodes receive the valid input $\left<1, v, \sigma \right>$, each correct node will output $1$.
    \item \rev{\textbf{Termination:} If each correct node receives an input, all correct nodes will eventually output.}
    \item \textbf{Shortcut $0$-output:} If all nodes receive the input $0$, each correct node can output after three communication rounds.
\end{itemize}

\vspace{-0.1cm}
\subsubsection{Construction of AABA}
We propose a construction of AABA, named $\Pi_{aaba}$, which leverages any existing ABA protocol in a black-box manner.
Specifically, $\Pi_{aaba}$ introduces a pre-processing component before running an existing ABA protocol.

\begin{algorithm}[!tp]
  \caption{Construction of AABA: $\Pi_{aaba}$ (for $p_i$)}
  \label{alg:aaba} 
  \setcounter{AlgoLine}{0}
  \SetAlgoNoEnd
  \SetKwProg{On}{on}{:}{}

  \textbf{Let} $I_i$ denote the input received by $p_i$; $S \leftarrow \emptyset$; $cnt_0 \leftarrow 0$\;
  \medskip

  \nonl // \texttt{amplification phase} \\
  \textbf{broadcast} $\left<\texttt{amp}, I_i\right>$\;
  \On{receiving $\left<\texttt{amp}, I_j \right>$ from $p_j$}{
      \If{$I_j$ is decoded as $\left<1, v, \sigma\right>$ \textbf{and} $Q(v, \sigma)$=\texttt{true}}{
        \If{$p_i$ has not broadcast a \texttt{sho1} message} {
            \textbf{broadcast} $\left<\texttt{sho1}, 1\right>$\;
        }
      } \ElseIf{$I_j$ is decoded as $0$}{
          $cnt_0 \leftarrow cnt_0 +1$\;
          \If{$cnt_0\text{=}n\text{-}f$ \textbf{and} $p_i$ has not broadcast \texttt{sho1} msg.}{
            \textbf{broadcast} $\left<\texttt{sho1}, 0\right>$\;
          }
      }
  }
  \medskip  

  \nonl // \texttt{shortcut phase} \\
  \On{receiving $\left<\texttt{sho1}, b \right>$ from $f+1$ nodes}{
    \If{$p_i$ has not broadcast $\left<\texttt{sho1}, b \right>$}{
        \textbf{broadcast} $\left<\texttt{sho1}, b\right>$\;
    }
  }
  \On{receiving $\left<\texttt{sho1}, b \right>$ from $n-f$ nodes}{
    $S \leftarrow S \bigcup \{b \}$ \;
    \If{$p_i$ has not broadcast a \texttt{sho2} message}{
        \textbf{broadcast} $\left<\texttt{sho2}, b\right>$\;
    }
  }
  \On{receiving $n-f$ \texttt{sho2} messages whose bits are in $S$ }{
    \If{all these messages contain $0$}{
        \textbf{output} $0$\;
    } 
    \If{at least one message contains $0$}{
        \textbf{input} $0$ to the subsequent ABA protocol
    } \Else {
        \textbf{input} $1$ to the subsequent ABA protocol
    }
  }
  \medskip

  \nonl // \texttt{output from ABA} \\
  \On{outputting $b$ from ABA}{
    \textbf{output} $b$ if not yet
  }
  
\end{algorithm}

\begin{figure}[!tp]
    \centering
    \vspace{-1.2cm}
    \includegraphics[width=\linewidth]{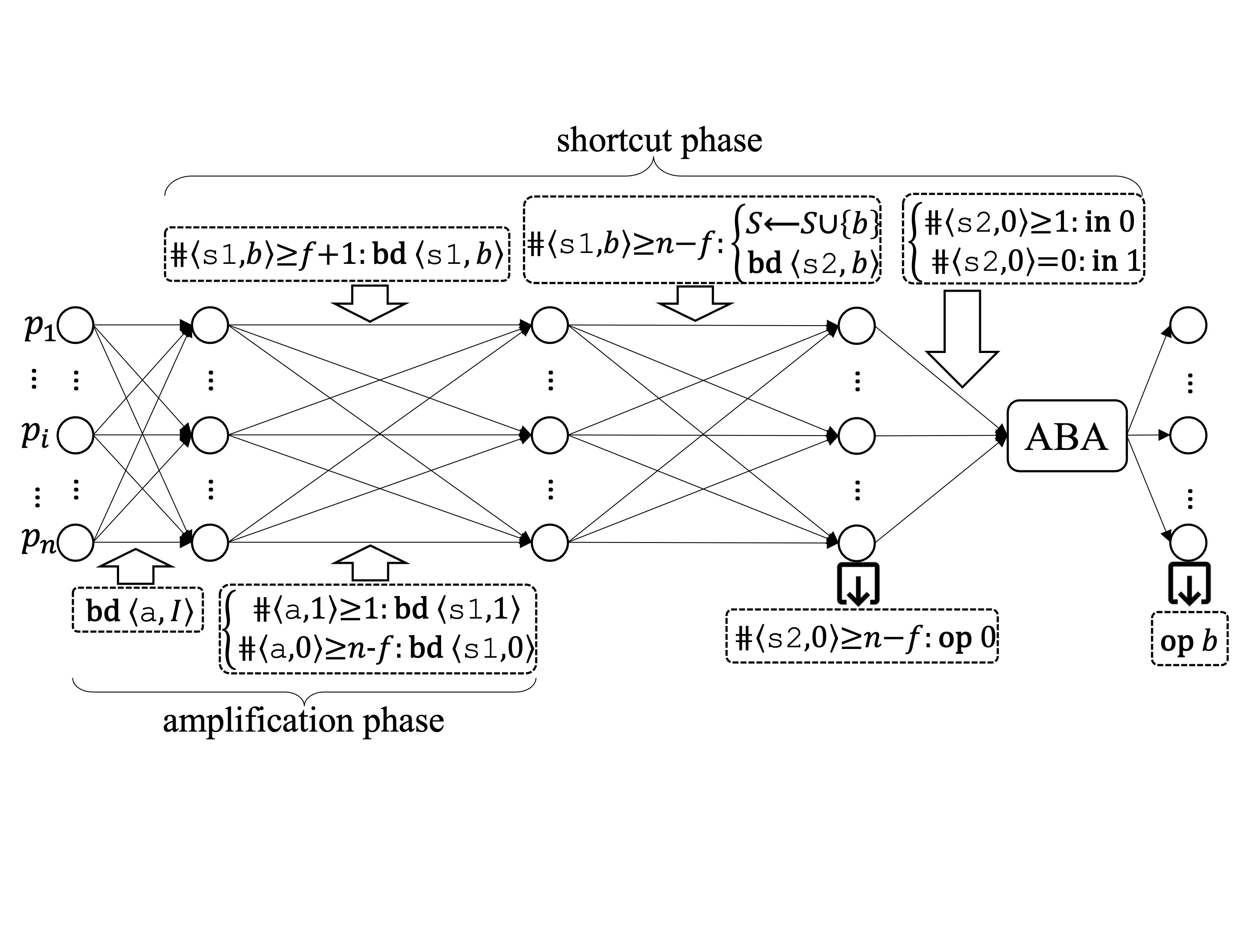}
    \vspace{-2.1cm}
    \caption{\rev{Schematic diagram of $\Pi_{aaba}$, where `bd,' `in,' and `op' denote the actions of `broadcast,' `input,' and `output.' Besides, `a,' `s1,' and `s2' represent the tags of `\texttt{amp},' `\texttt{sho1},' and `\texttt{sho2}.'}}
    \vspace{-0.5cm}
    \label{fig:alg1-fig}
\end{figure}

The pre-processing component of $\Pi_{aaba}$ is designed in two phases, with the first amplifying the input of $1$, named the amplification phase, and the second facilitating shortcut output of $0$, named the shortcut phase.
\rev{The pseudocode and schematic diagram of $\Pi_{aaba}$ are presented as Algorithm~\ref{alg:aaba} and Figure~\ref{fig:alg1-fig}, respectively.
In the amplification phase, each node broadcasts its input in the \texttt{amp} message (Line 2 of Algorithm~\ref{alg:aaba} and $\left<\texttt{a}, I\right>$ messages in Figure~\ref{fig:alg1-fig}) and waits to receive at least $n-f$ \texttt{amp} messages.
If a node receives at least one \texttt{amp} message with the valid bit of $1$, it will broadcast the \texttt{sho1} message of $1$ for the shortcut phase (Lines 4-6 and $\left<\texttt{s1}, 1\right>$ in Figure~\ref{fig:alg1-fig}).
Conversely, when it receives $n-f$ \texttt{amp} messages with $0$, it will broadcast the \texttt{sho1} message of $0$ (Lines 8-10 and $\left<\texttt{s1}, 0\right>$ in Figure~\ref{fig:alg1-fig}).
Through the amplification phase, as long as $f+1$ correct nodes have the initial input of $1$, all correct nodes will input $1$ to the subsequent shortcut phase, thereby amplifying the input of $1$.}

The shortcut phase consists of two steps, with the first step mirroring a filter functionality designed by MMR-ABA~\cite{mostefaoui2014signature}.
Within the first step, if a node receives $b$ from at least $f+1$ nodes and it has not yet broadcast a \texttt{sho1} message containing $b$, it will also broadcast a $\left<\texttt{sho1}, b\right>$ (Lines 11-13), even though it has broadcast $1-b$ before.
In the meanwhile, once a node receives $\left<\texttt{sho1}, b\right>$ from at least $n-f$ nodes, it will add $b$ to a local set $S$. \rev{If it has not broadcast a \texttt{sho2} message, it will broadcast one containing $b$ (Lines 14-17 and $\left<\texttt{s2}, b\right>$ messages in Figure~\ref{fig:alg1-fig}).}
Intuitively, this step (Lines 11-17) guarantees that a correct node will not include $b$ in its set $S$ if $b$ is input to the shortcut phase only by Byzantine nodes. 

In the second step, the node will wait to receive $n-f$ \texttt{sho2} messages, each of which must contain a bit in the set $S$.
If all these \texttt{sho2} messages contain $0$, the node will output $0$ directly (Lines 19-20).
If at least one \texttt{sho2} message contains $0$, the node will input $0$ to the subsequent ABA protocol (Lines 21-22); otherwise, it will input $1$ (Lines 23-24). It will then take the output from ABA as its output from $\Pi_{aaba}$ (Lines 25-26).

If all nodes input $0$ to $\Pi_{aaba}$, each correct node can output $0$ through the shortcut mechanism at Line 20, resulting in a latency of three communication rounds. 
\rev{Note that the shortcut mechanism cannot be exploited by the adversary to undermine the protocol’s correctness. In simple terms, when a correct node outputs via the shortcut mechanism, it must have received $n-f$ \texttt{sho2} messages containing $0$ (Line 19). Since $n \geq 3f+1$, each correct node will receive at least one \texttt{sho2} message with $0$ and then input $0$ into the ABA protocol (Lines 21-22). Due to ABA's validity property, it will output $0$, ensuring consistency with the shortcut output.}
Detailed correctness analysis of $\Pi_{aaba}$ are given in \autoref{sec:anal_aaba}.

\begin{algorithm}[!tp]
  \caption{Early-stopping mechanism for $\Pi_{aaba}$ (for $p_i$)}
  \label{alg:early-stop} 
  \setcounter{AlgoLine}{0}
  \SetAlgoNoEnd
  \SetKwProg{On}{on}{:}{}

  \nonl // \texttt{after outputting $0$ in Line 20 of Algorithm~\ref{alg:aaba}} \\
  \textbf{broadcast} $\left<\texttt{stop}, 0\right>$\;

  \On{receiving $f+1$ \texttt{stop} messages containing $0$ }{
    \If{$p_i$ has not broadcast a \texttt{stop} message yet} {
        \textbf{broadcast} $\left<\texttt{stop}, 0\right>$\;
    }
    \If{$p_i$ has not outputted yet} {
        \textbf{output} $0$\;
    }
  }

  \On{receiving $n-f$ \texttt{stop} messages containing $0$ }{
    \textbf{exit} from current $\Pi_{aaba}$ instance
  }
\end{algorithm}

\textbf{Early-stopping mechanism.}
In $\Pi_{aaba}$, a correct node must continue to execute the subsequent ABA even if it has outputted $0$ (Line 20 of Algorithm~\ref{alg:aaba}).
This is necessary because the ABA protocol promises to terminate only if each correct node receives an input.
To address this, we introduce an early-stopping mechanism, described in Algorithm~\ref{alg:early-stop}, which enables a node to early exit without finishing the subsequent ABA protocol if $f+1$ correct nodes output $0$ at Line 20.
To be more specific, a node will broadcast a \texttt{stop} message after outputting $0$ in Line 20 of Algorithm~\ref{alg:aaba}.
Once receiving $f+1$ \texttt{stop} messages, a node will also broadcast a \texttt{stop} message containing $0$ (Lines 2-4 of Algorithm~\ref{alg:early-stop}). Furthermore, it will output $0$ if it has not done yet (Lines 5-6).
Once receiving $n-f$ \texttt{stop} messages, a node can exit from the current $\Pi_{aaba}$ instance. Particularly, it will stop its participation in the subsequent ABA protocol.
The correctness of the early-stopping mechanism is analyzed in Appendix~\ref{sec:anal-early-stop}.

\begin{algorithm}[!tp]
  \caption{\proto~protocol (for $p_i$)}
  \label{alg:proto} 
  \setcounter{AlgoLine}{0}
  \SetAlgoNoEnd
  \SetKwProg{On}{on}{:}{}
  \SetKwProg{cWhile}{while}{:}{}
  \SetKwProg{Define}{define}{:}{}

  \textbf{Let} $\acsq^k.v$ denote the variable $v$ in $\acsq^k$;  $k \leftarrow 1$\;

  \cWhile{\texttt{true}}{
    \If{$\acsq^k$ has not been activated}{
      \textbf{activate} $\acsq^k$\;
    }
    \On{$\left|\acsq^k.M_2\right| = n-f$}{
      \textbf{activate} $\acsq^{k+1}$\;
    }
    \On{$\left|\acsq^{k+1}.M_2\right| \geq 1$}{
      \rev{\textbf{activate} $\acsq^k.trigger$}\;
    }
    \textbf{wait until} $\acsq^k$ returns\;
    $k \leftarrow k+1$\;
  }

\end{algorithm}

\begin{algorithm}[!tp]
  \caption{$\Pi_{acsq}$ with the instance identity $k$ (for $p_i$)}
  \label{alg:acsq} 
  \setcounter{AlgoLine}{0}
  \SetAlgoNoEnd
  \SetKwProg{On}{on}{:}{}
  \SetKwProg{cWhile}{while}{:}{}
  \SetKwProg{Define}{define}{:}{}

  \textbf{Let} $B_i$ denote the block proposed by $p_i$\;
  $M_{1} \leftarrow []$; $M_{2} \leftarrow []$; $M_{acs} \leftarrow []$; $S_{a} \leftarrow \emptyset$; $idx \leftarrow 0$; $S_{ex} \leftarrow \emptyset$\;
  \medskip
    
  \nonl // \texttt{broadcast stage} \\
  \textbf{broadcast} $B_i$ through $\gbc_i$\;
  \On{delivering $\left<B_j, 1, \sigma 1\right>$ from $\gbc_j$}{
    $M_{1}[j] \leftarrow \left<B_j, 1, \sigma 1\right>$\;
  }
  
  \On{delivering $\left<B_j, 2, \sigma 2\right>$ from $\gbc_j$}{
    $M_{2}[j] \leftarrow \left<B_j, 2, \sigma 2\right> $\;
    $M_{acs}[j] \leftarrow B_j $; // \texttt{include it to the \acs~set directly} \\
    $PartialSort(k, M_{acs}, S_{ex}, j)$;  \\ 
  }

  \textbf{wait until} $\left| M_{2} \right| = n$ \textbf{or} the \rev{$trigger$ activation}\;
  \If{\rev{the trigger is activated}}{
    \textbf{wait until} $\left| M_{2} \right| \geq n-f $\;
    \textbf{stop} sending partial signatures in the broadcast stage\;
  }

  \medskip
  \nonl // \texttt{agreement stage} \\
  \ForEach{$j \in [1..n]$ s.t. $M_{2}[j] = \bot $}{
    $S_a \leftarrow S_a \cup \left\{j\right\}$\;
    \If{$M_{1}[j] \neq \bot$}{
      $(B, g, \sigma ) \leftarrow M_{1}[j]$\;
      \textbf{input} $\left<1, B.d, \sigma \right>$ to $\aaba_j$; //\texttt{$B.d$ denotes $B$'s digest} \\
    } \Else{
      \textbf{input} $0$ to $\aaba_j$\;
    }
  }

  \On{$\aaba_j$ outputs $b$}{
      \If{$b=1$}{
          $M_{acs}[j] \leftarrow B_j$; // \texttt{include it to the \acs~set} \\
      } \Else{
         $S_{ex} \leftarrow S_{ex} \bigcup \left\{j\right\}$\;
      }
      $PartialSort(k, M_{acs}, S_{ex}, j)$; \\
  }

  \medskip
  \nonl // \texttt{delivery-assistance mechanism} \\
  \On{receiving an $\aaba_j$ message from $p_t$} {
      \If{$M_2[j] \neq \bot$}{
        \textbf{send} $M_2[j]$ to $p_t$\;
      }
  }
  \On{receiving $\left<B_j, 2, \sigma 2\right>$}{
      $M_{2}[j] \leftarrow \left<B_j, 2, \sigma 2\right>$\;
      $M_{acs}[j] \leftarrow B_j $; // include it to the \acs~set \\
      $PartialSort(k, M_{acs}, S_{ex}, j)$; \\
      \textbf{stop} participating in $\aaba_j$\;
  }
  
  \medskip
  \textbf{wait until} $\aaba_j$ outputs \textbf{or} stops through \textit{Line 34},  $\forall j \in S_a$\;
  \textbf{return} from current $\acsq$ instance
\end{algorithm}

\section{\proto~Design}

\subsection{Overall design}
\begin{figure}
    \centering
    \vspace{-0.4cm}
    \includegraphics[width=\linewidth]{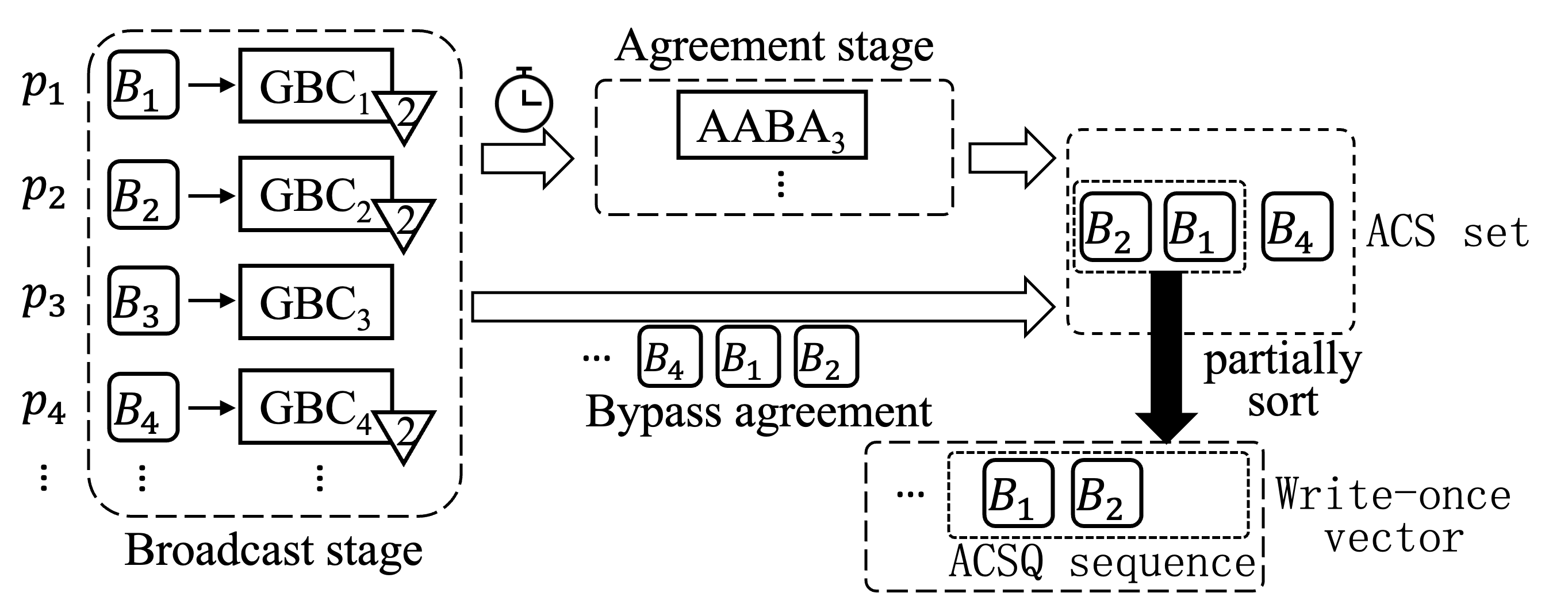}
    \vspace{-0.8cm}
    \caption{$\Pi_{acsq}$ with the partial sorting mechanism}
    \vspace{-0.2cm}
    \label{fig:acsq-struc}
\end{figure}

Like in other existing BFT consensus constructed based on the ACSQ protocol, \proto~also advances by executing successive $\acsq$ instances, which is described in Algorithm~\ref{alg:proto}.
The ACSQ protocol also comprises two parts: the ACS protocol and the block partial-sorting mechanism, as described in Algorithm~\ref{alg:acsq}.
The ACS protocol consists of the broadcast stage and the agreement stage.
A schematic diagram of $\Pi_{acsq}$ is illustrated in Figure~\ref{fig:acsq-struc}.

\subsubsection{Broadcast stage}
Within this stage, each node broadcasts its block through the GBC protocol (Lines 3-8).
If the node $p_i$ delivers the block $B_j$ with the grade $2$, $p_i$ can directly include $B_j$ into its \acs, without running the \aaba~instance for $B_j$, as shown by $B_2$, $B_1$, and $B_4$ in Figure~\ref{fig:acsq-struc}.
The node waits for all blocks in this ACSQ instance to be grade-$2$ delivered or \rev{until the agreement trigger is activated, as defined in Section~\ref{sec: trigger}.}
If it is the latter, the node will continue to wait until at least $n-f$ blocks have been grade-$2$ delivered if not yet.
After that, the node stops generating or sending the partial threshold signatures in the broadcast stage (Lines 10-13).
If all blocks are grade-$2$ delivered before \rev{the trigger activation}, node $p_i$ can skip the agreement stage, thus reducing latency and addressing \textbf{issue 1} outlined in Section~\ref{sec:cons_async_proto}.

\begin{figure}
    \centering
    \includegraphics[width=0.95\linewidth]{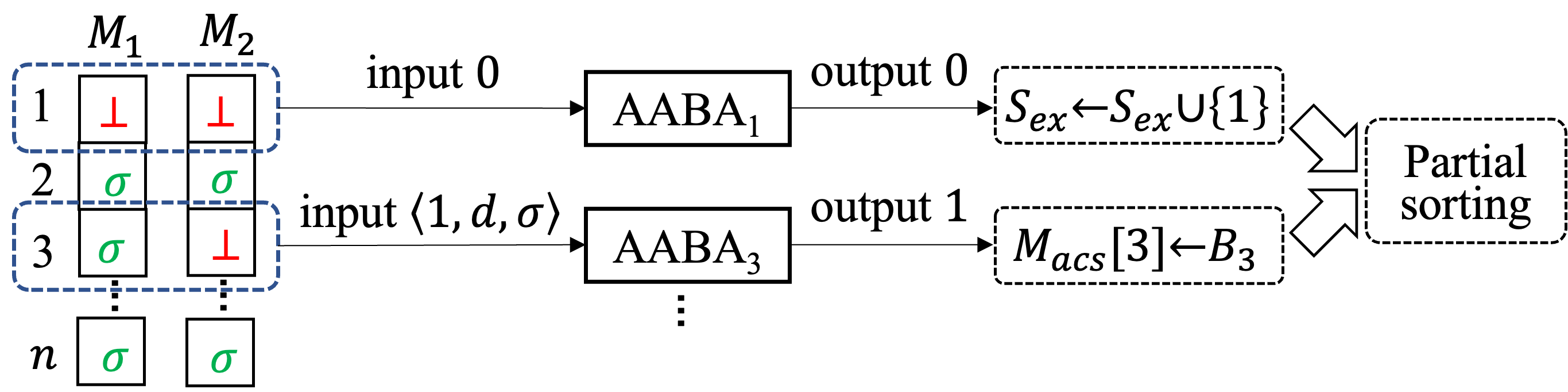}
    \vspace{-0.3cm}
    \caption{\rev{Schematic diagram of the agreement stage}}
    \vspace{-0.3cm}
    \label{fig:agree-fig}
\end{figure}

\vspace{-0.1cm}
\subsubsection{Agreement stage}\label{sec:agree-stage}
\rev{$p_i$ will execute an \aaba~instance for each block $B_j$ that has not been grade-$2$ delivered, as described in Figure~\ref{fig:agree-fig}}.
Specifically, $p_i$ constructs an input to \aaba~instance $\aaba_j$ based on its delivery situation of $B_j$.
\rev{If it has delivered $B_j$ with the grade $1$ (e.g., $B_3$ in Figure~\ref{fig:agree-fig}), it will input $1, d, \sigma 1$ to $\aaba_j$, where $d$ denotes the digest of $B_j$.
Otherwise, if $B_j$ has not been delivered at all (e.g., $B_1$ in Figure~\ref{fig:agree-fig}), it will input $0$ to $\aaba_j$ (Lines 14-20).}


$p_i$ will wait for each $\aaba$ instance to terminate, either eventually outputting or delivering the corresponding blocks through the delivery-assistance mechanism.
The delivery-assistance mechanism will be detailed in Section~\ref{sec:deliv-assist}.
\rev{If $\aaba_j$ outputs $1$ (e.g., $\aaba_3$ in Figure~\ref{fig:agree-fig}), the node will also include $B_j$ to the \acs~set (Lines 21-23).
Otherwise, such as $\aaba_1$ in Figure~\ref{fig:agree-fig}, it will be marked in a set $S_{ex}$ (Lines 24-25), which will be later accessed during the partial sorting process.}
Due to the conciseness, we omit a trivial query process after receiving $1$ from $\aaba_j$ (Line 22) and before including $B_j$ to the \acs~set (Line 23).
Specifically, $p_i$ may need to query $B_j$ since it may not have received $B_j$.
This can be done by broadcasting a query request containing the digest of $B_j$.
A correct node will respond with the data of $B_j$ after receiving this request.
We will prove in Lemma 1 of Section~\ref{sec:anal-pi-acsq} that at least one correct node must have received $B_j$ and can complete the response.

\begin{algorithm}[!tp]
  \caption{Partially sorting protocol (for $p_i$)}
  \label{alg:partial-sort} 
  \setcounter{AlgoLine}{0}
  \SetAlgoNoEnd
  \SetKwProg{cWhile}{while}{:}{}
  \SetKwProg{Define}{define}{:}{}

  $doneACSQId \leftarrow 0$\; 

  \Define{$PartialSort(k, M, S, idx)$}{
    \textbf{wait until} $doneACSQId=k-1$\;
    \cWhile{$idx < n$ \textbf{and} ($M[idx+1] \neq \bot$ \textbf{or} $idx+1 \in S$)}{
      \If{$M[idx+1] \neq \bot$}{
        $l \leftarrow \left| \mathcal{C}_i \right|$; $\mathcal{C}_i[l+1] \leftarrow M[idx+1]$\;
      }
      $idx \leftarrow idx+1$\;
    }
    \If{$idx=n$}{
      $doneACSQId=k$
    }
    \textbf{return} $idx$\;
  }
\end{algorithm}

\subsubsection{Partial sorting}\label{sec:par-sort}
A node in \proto~can sort blocks without waiting for all blocks to have been decided, which is named a partial sorting mechanism.
Anytime a new block is decided as included or excluded in the \acs~set, the node can call the partial sorting function to commit blocks, as shown in Line 9, Line 26, and Line 33 of Algorithm~\ref{alg:acsq}.
The partial sorting function is described in Algorithm~\ref{alg:partial-sort}, which accepts four parameters: $k$ denotes the identity of the $\acsq$~instance that calls the function, $M$ represents the set containing all blocks decided as included, $S$ represents the set marking all blocks decided as excluded, and $idx$ denotes the last index of $\acsq^k$ processed by the sorting protocol.

An index can be processed if all smaller indices have been processed and a block with this index has been decided. 
Furthermore, if a block with this index is included in the \acs~set, this block can be sorted immediately.
A one-larger index $idx+1$ will be processed after another $idx$ until all indices have been processed or a block with $idx+1$ has not been decided (Lines 4-7). 
The partial-sorting mechanism allows blocks to be committed continuously, eliminating the need to wait for the slowest block to be decided. 
\rev{This not only improves latency stability, addressing \textbf{issue 2} from Section~\ref{sec:cons_async_proto}, but also reduces overall latency.}
As a practice common in the asynchronous BFT protocol and having been introduced in Section~\ref{sec:smr-acsq}, blocks in an $\acsq$ instance can be committed only after all $\acsq$ instances with smaller identities have been processed, which is ensured by Line 3 and Lines 8-9 of Algorithm~\ref{alg:partial-sort}.

\vspace{-0.2cm}
\subsection{\rev{Agreement trigger}}\label{sec: trigger}
At a high level, the \rev{agreement trigger} in an $\acsq$ instance is designed based on an event in the subsequent $\acsq$ instance.
To be more specific, if a node grade-$2$ delivers $n-f$ blocks in an instance $\acsq^{k}$, it will activate the next instance $\acsq^{k+1}$, as shown by Lines 5-6 of Algorithm~\ref{alg:proto}.
If some block in $\acsq^{k+1}$ has been grade-$2$ delivered, \rev{the trigger in $\acsq^{k}$ is activated} (Lines 7-8 of Algorithm~\ref{alg:proto}).

If it is in a favorable situation, all blocks in $\acsq^{k}$ can be grade-$2$ delivered before any block in $\acsq^{k+1}$ is grade-$2$ delivered, and \rev{the trigger will not be activated at all}.
Additionally, under this favorable situation, all blocks can be committed, leading to higher throughput and resolving \textbf{issue 3} outlined in Section~\ref{sec:cons_async_proto}.

\vspace{-0.2cm}
\subsection{Delivery-assistance mechanism}\label{sec:deliv-assist}
As described in Section~\ref{sec:agree-stage}, a node will not activate the $\aaba_j$ instance for a block $B_j$ if it has grade-$2$ delivered $B_j$.
Other nodes that activate the $\aaba_j$ instance may not output from $\aaba_j$, since the AABA protocol only promises termination if each correct node receives an input.
To address this, we introduce a delivery-assistance mechanism, as described by Lines 27-34 of Algorithm~\ref{alg:acsq}.
Specifically, if a correct node $p_i$ receives an $\aaba_j$ message from $p_t$ and $p_i$ has grade-$2$ delivered $B_j$, $p_i$ will send $\left<B_j, 2, \sigma 2\right>$ to $p_t$ (Lines 27-29).
After receiving $\left<B_j, 2, \sigma 2\right>$, $p_t$ deal with $B_j$ as if $B_j$ is grade-$2$ delivered through the $\gbc_j$ instance.
Besides, $p_t$ will stop its participation in $\aaba_j$ (Lines 30-34).

\vspace{-0.2cm}
\subsection{Performance analysis}
We conduct the performance analysis by adopting the $\Pi_{aaba}$ protocol in the agreement stage.
Besides, the ABA protocol absorbed in $\Pi_{aaba}$ is implemented using the ABY-ABA protocol~\cite{abraham2022efficient}, which has an expected worst latency of 9 communication rounds.
Since it is hard to perform a quantitative analysis of the sorting process, our primary focus is on analyzing the latency of generating the \acs~set within an $\acsq$ instance, in terms of the communication rounds.

In a favorable situation where all nodes are correct, and the network condition is well, all blocks can be grade-$2$ delivered and included into the \acs~set before the \rev{trigger activation}, resulting in a latency of 3 communication rounds to generate the \acs~set.

In a less favorable situation where nodes can only crash before the $\acsq$ instance activates, some blocks fail to be grade-$2$ delivered before the \rev{trigger activation}. However, all correct nodes will input $0$ to these \aaba~instances, which takes 3 communication rounds to output through the shortcut outputting mechanism.
Since the \rev{trigger is activated} by grade-$2$ delivering a block in the next $\acsq$ instance and the next $\acsq$ instance is activated after grade-$2$ delivering $n-f$ blocks in the current $\acsq$ instance, the time for the \rev{trigger activation} equals the sum of two \gbc~instances, namely 6 communication rounds.
Therefore, the latency to generate the \acs~set in this situation is 9 communication rounds.

In the worst situation, $f$ \aaba~instances are activated and cannot output through the outputting mechanism. It takes 12 rounds in expectation to output from \aaba.
Therefore, the latency to generate the \acs~set in this situation is $6+12\cdot log(n)$ rounds.

\section{Correctness Analysis}\label{sec: correc-anal}
\subsection{Analysis on $\acsq$}\label{sec:anal-pi-acsq}
In this section, we prove that $\acsq$ implements an ACSQ protocol.
\rev{We start with an intuitive analysis, followed by a rigorous one.}

\begin{revpara}
\subsubsection{Intuitional analysis}
This analysis involves an assessment of $\acsq$'s resilience against three common types of attacks.
\vspace{-0.1cm}
\paragraph{Attack 1: send inconsistent blocks} In the broadcast stage, a Byzantine node may intentionally send inconsistent blocks in an attempt to make nodes deliver conflicting blocks. However, in the GBC protocol, correct nodes must collect voting messages (i.e., partial threshold signatures) for the same block from at least $n-f$ nodes before delivering that block. Moreover, each correct node will vote for only one block in a given GBC instance. Since $n \geq 3f+1$, at most one block can be voted on by $n-f$ nodes. In other words, if two correct nodes deliver blocks $B$ and $B'$, then $B$ and $B'$ must be identical, effectively thwarting the attack.

\vspace{-0.15cm}
\paragraph{Attack 2: delay messages} In a GBC instance with a correct broadcaster, the adversary can delay messages to cause only a subset of correct nodes to deliver the block with grade $2$, adding it to the \acs\ set, while others may either deliver the block with grade $1$ or not at all. The goal is to cause discrepancies between \acs\ sets, compromising safety. However, GBC's delivery-correlation property ensures that, in this case, at least $f+1$ correct nodes deliver the block with grade $1$, which will then input $1$ in AABA. AABA's biased-validity property guarantees that once $f+1$ correct nodes input $1$, all nodes will output $1$, adding the block to the \acs\ set. This ensures the consistency of \acs\ sets across all nodes, protecting against the attack.

\vspace{-0.15cm}
\paragraph{Attack 3: input wrong bits to AABA} In AABA, a Byzantine node may intentionally input a wrong bit to induce unreasonable outputs, in two ways: (1) The Byzantine node aids a correct node in delivering a block with grade $2$, adding it to the \acs\ set, but then inputs $0$ to AABA, hoping others to exclude the block. (2) The Byzantine node does not deliver the block but inputs $1$ to AABA. In the first case, GBC's delivery-correlation property ensures at least $f+1$ correct nodes input $1$ to AABA. Combined with AABA's biased-validity property, this guarantees AABA outputs $1$, regardless of the Byzantine input. In the second case, AABA requires a proof $\sigma$ for inputting $1$, preventing the Byzantine node from inputting an invalid $1$. Thus, $\acsq$ effectively defends against attacks involving incorrect bit inputs to AABA.
\end{revpara}

\subsubsection{Rigorous analysis}\label{sec:rig-anal}
\rev{The rigorous analysis focuses on proving whether $\acsq$ satisfies various ACSQ properties defined in Section~\ref{sec:acsq}.}
\smallskip

\textsc{Lemma 1.} \textit{Within a $\acsq$ instance, if a correct node includes $B$ in its \acs~set, then all correct nodes will include $B$ in their \acs~sets.}
\vspace{-0.2cm}
\begin{proof}
    As shown in Algorithm~\ref{alg:acsq}, a block will be decided at three possible points: at the end of \gbc~instance (Line 8), at the end of the \aaba~instance (Line 23), and through the delivery-assistance mechanism (Line 32).
    \rev{We refer to the decision at these three points as \gbc-decide, \aaba-decide, and \texttt{DA}-decide, respectively, for short.} Consider the following two cases:
    
    \rev{\textbf{Case 1: At least one correct node \gbc-decides to include $B$.}} Denote this node as $p_i$. According to GBC's delivery-correlation property, at least $f+1$ correct nodes must have delivered $B$ with grade $1$.
    Each of these nodes will input $1$ to \aaba~if it has not grade-$2$ delivered $B$.
    On the other hand, for another correct node $p_j$, \rev{if $p_j$ \gbc-decides to include a block, according to GBC's consistency property, $B'$ must be identical to $B$. If $p_j$ \aaba-decides a block, according to AABA's biased-validity property, the \aaba~instance must output $1$.} According to AABA's $1$-validity property, some node must have inputted $\left<1, v, \sigma \right>$ s.t. $Q(v, \sigma)$=\texttt{true}. Since $\sigma$ is a proof of grade-$1$ delivery in GBC, according to the receipt-correlation property, at least $f+1$ correct nodes have received $B$. $p_j$ can acquire $B$ by broadcasting a query request of $B$ if it has not received $B$ from the broadcaster. In other words, $p_j$ can definitely include $B$ in the \acs~set.    
    \rev{If $p_j$ has not \gbc-decided or \aaba-decided a block, it will eventually receive from $p_i$ a delivery-assistance message (Line 30 of Algorithm~\ref{alg:acsq}), namely \texttt{DA}-deciding to include $B$.}
    
    \rev{\textbf{Case 2: No correct node \gbc-decides to include $B$.}} In this case, \rev{all nodes can only \aaba-decide on a block}. If a correct node includes $B$ in the \acs~set, it must output $1$ from the corresponding \aaba~instance. 
    According to AABA's agreement property, each correct node will output $1$ from this \aaba~instance.
    According to AABA's $1$-validity property, some node must have inputted $\left<1, v, \sigma \right>$ s.t. $Q(v, \sigma)$=\texttt{true}. Similar to the analysis in Case 1, each correct node can eventually receive $B$ and include $B$ in the \acs~set.
\end{proof}

\begin{revpara}
\textsc{Theorem 2 (Agreement).} \textit{Within a $\acsq$ instance, if a correct node outputs $q$, then every correct node outputs $q$.}
\vspace{-0.2cm}
\begin{proof}
    If a correct node excludes a block from the \acs~set, it must output $0$ from \aaba. By AABA's agreement property, each correct node will output $0$ from \aaba~and exclude the block from its \acs~set.
    Conversely, if a correct node includes a block into the \acs~set, then, by Lemma 1, each correct node will include this block.
    When blocks in the \acs~set are sorted by their block numbers, the resulting sequences will also match, ensuring agreement.
\end{proof}
\end{revpara}
\medskip

\textsc{Theorem 3 (Validity).} \textit{Within a $\acsq$ instance, if a correct node outputs a sequence $q$, then $\left| q \right| \geq n-f$.}
\vspace{-0.2cm}
\begin{proof}
    As described by Line 10 and Line 12 of Algorithm~\ref{alg:acsq}, a correct node will wait to deliver at least $n-f$ blocks with grade $2$, each of which will be included in the \acs~set and the $\acsq$ sequence.
    Thus, the sequence will contain at least $n-f$ elements.
\end{proof}
\medskip

\textsc{Theorem 4 (Totality).} \textit{Within a $\acsq$ instance, \rev{if each correct node receives an input, all correct nodes will eventually output}.}
\vspace{-0.2cm}
\begin{proof}
    Since there are at least $n-f$ correct nodes, each correct node can deliver at least $n-f$ blocks with grade $2$ and then finish the broadcast stage in Algorithm~\ref{alg:acsq}.
    Next, we prove that for each block without being decided in the broadcast stage, a node can decide it during the agreement stage.

    Assume a node $p_i$ does not decide a block $B$ in the broadcast stage. We consider the following two cases. First, if no correct node decides $B$ in the broadcast stage, all correct nodes will activate an \aaba~instance for $B$. According to AABA's termination property, $p_i$ will output from AABA and decide on $B$. Second, if some correct node decides $B$ in the broadcast stage, it will send a delivery-assistance message to $p_i$. Therefore, $p_i$ can decide on $B$ after receiving this delivery-assistance message if it has not made a decision based on AABA's output.
    
    Thus, each correct node can eventually output from $\Pi_{acsq}$.
\end{proof}

\medskip
\textsc{Theorem 5 (Optimistic validity).} \textit{Within a $\acsq$ instance, if all nodes are correct and the network conditions are favorable, each correct node can output a sequence comprising inputs from all nodes, such that $\left| q \right| = n$.}
\vspace{-0.2cm}
\begin{proof}
    If all nodes are correct and the network conditions are favorable, each correct node can grade-$2$ deliver all blocks before the \rev{trigger activation} during the broadcast stage.
    Therefore, the outputted sequence will be in size of $n$.
\end{proof}

\subsection{Analysis on \proto}\label{sec:anal-proto}
In this section, we prove that our \proto~protocol constructed based on $\acsq$ correctly implements BFT.

\medskip
\textsc{Theorem 6 (Safety).} \textit{For two nodes $p_i$ and $p_j$, if $\mathcal{C}_i[r] \neq \bot$ and $\mathcal{C}_j[r] \neq \bot$, then $\mathcal{C}_i[r] = \mathcal{C}_j[r]$.}
\vspace{-0.2cm}
\begin{proof}
    As described in Section~\ref{sec:par-sort}, blocks in the $\Pi_{acsq}^h$ instance can be committed only if all $\Pi_{acsq}^m$ ($m<h$) instances have been processed.
    For ease of presentation, we denote the $\acsq$ sequence generated by node $p_i$ through $\Pi_{acsq}^m$ as $\mathcal{A}_i^m$. Additionally, the vector updated by node $p_i$ after completing the $\Pi_{acsq}^m$ instance is denoted as $\mathcal{C}_i^m$. According to Theorem 2, $\mathcal{A}_i^m$ must be equal to $\mathcal{A}_j^m$, and thus $\mathcal{C}_i^m$ must also be equal to $\mathcal{C}_j^m$.

    Without loss of generality, we assume that $\mathcal{C}_i[r]$ and $\mathcal{C}_j[r]$ are committed by $p_i$ and $p_j$ in $\mathcal{A}_i^{h1}$ and $\mathcal{A}_j^{h2}$, respectively.
    If $h1=1$, then $h2$ must also equal $1$. In other words, $\mathcal{C}_i[r]$ and $\mathcal{C}_j[r]$ are committed in the first $\acsq$ instance.
    According to Theorem 2, $\mathcal{C}_i[r]$ and $\mathcal{C}_j[r]$ must be identical.

    If $h1 > 1$, then $h2$ must be greater than $1$, and moreover, $h1$ and $h2$ will be equal.
    Thus, $\mathcal{C}_i^{h1-1}$ must equal to $\mathcal{C}_i^{h2-1}$. Denote the lengths of $\mathcal{C}_i^{h1-1}$ and $\mathcal{C}_j^{h2-1}$ as $l_i$ and $l_j$, respectively.
    $l_i$ and $l_j$ will also be identical.
    Thus, $\mathcal{C}_i[r]$ is committed by $p_i$ in $\Pi_{acsq}^{h1}$ with index $r-l_i$, and $\mathcal{C}_j[r]$ is committed by $p_j$ in $\Pi_{acsq}^{h2}$ with index $r-l_j$.
    Since $h1=h2$ and $l_i=l_j$, $\mathcal{C}_i[r]$ and $\mathcal{C}_j[r]$ must be identical.
\end{proof}

\textsc{Theorem 7 (Liveness).} \textit{If a transaction $\tx$ is included in every correct node's buffer, each one will eventually commit $\tx$.}
\vspace{-0.2cm}
\begin{proof}
    According to Theorem 3 and Theorem 4, in a $\acsq$ instance, each correct node will output a sequence $q$ where $\left| q \right| \geq n-f$.
    In this sequence, at least $n-2f$ blocks are proposed by correct nodes.
    When $\tx$ is present in every correct node's buffer, each one will include $\tx$ in its newly proposed block if $\tx$ has not already been committed.
    Therefore, if $\tx$ remains uncommitted, the sequence $q$ must include blocks containing $\tx$, which will be committed in this $\acsq$ instance.
    Thus, \proto~guarantees the liveness property.
\end{proof}

\section{Implementation and Evaluation}
\subsection{Implementation \& settings}\label{sec:impl-set}
To evaluate \proto's performance, we develop a prototype system implementation. \rev{We select HBBFT~\cite{miller2016honey}, Dumbo\footnote{We utilize sMVBA~\cite{guo2022speeding} as the MVBA protocol of Dumbo to achieve good performance.}~\cite{guo2020dumbo}, MyTumbler (abbreviated as MyTblr)~\cite{liu2023flexible}, and Narwhal\&Tusk (abbreviated as Tusk)~\cite{danezis2022narwhal} as baselines. HBBFT and MyTblr represent the BKR paradigm, Dumbo embodies CKPS, and Tusk represents a state-of-the-art \textit{Directed Acyclic Graph} (DAG)-based protocol. All implementations---HBBFT, Dumbo, MyTblr, and \proto---are coded in Rust within the same code framework, without utilizing threshold encryption for transactions~\cite{miller2016honey}.}
\rev{Specifically, we use ed25519-dalek\footnote{https://github.com/dalek-cryptography/ed25519-dalek} for elliptic curve based signatures and threshold\_crypto\footnote{https://github.com/poanetwork/threshold\_crypto} for threshold signatures. We choose the MMR version of the ABA protocol~\cite{mostefaoui2014signature} for implementation.}
\rev{For Tusk, we directly utilize the open-source codebase\footnote{https://github.com/facebookresearch/narwhal} directly.}
The experiments are conducted on AWS, with each node deployed on an m5d.2xlarge instance. These instances are equipped with 8 vCPUs and 32GB of memory and run Ubuntu 20.04, with a maximum network bandwidth of 10Gbps. To simulate a decentralized deployment, nodes are distributed across five regions globally: N. Virginia, Stockholm, Tokyo, Sydney, and N. California.
\rev{It is important to note that the signature algorithm has a minimal impact on the overall protocol performance. Therefore, in line with common practice in consensus evaluations, this section omits a detailed analysis of these signature libraries.}

To minimize the impact of experimental errors, each experiment is repeated three times, and we plot the average or error bars for each data point in the experiment.
We consider two situations: favorable and unfavorable. The favorable situation refers to a system without faulty nodes, while the unfavorable situation includes some faulty nodes. As in most mainstream consensus evaluations~\cite{blum2023abraxas, dai2023parbft, gelashvili2022jolteon}, we simulate faulty nodes in the system using crash failures.

Our prototype implementation is built on a mempool that broadcasts transactions via payloads, a method widely adopted by various blockchain systems~\cite{gelashvili2022jolteon, gelashvili2023block, blum2023abraxas, dai2023parbft}. 
\rev{In essence, a payload refers to a package that bundles multiple transactions together.} Each payload is configured to be 500 KB.
\rev{Besides, as in many other consensus studies~\cite{blum2023abraxas, dai2023parbft, miller2016honey, guo2020dumbo}, we generate mock transactions for evaluation.}
Each transaction is set to 512 bytes, with each block referencing up to 32 payloads.
\rev{These parameters are also commonly used in recent BFT evaluation works~\cite{gelashvili2022jolteon, dai2023parbft, danezis2022narwhal, shrestha2024sailfish}.}
\rev{Due to the small size of the blocks, we did not utilize erasure codes~\cite{cachin2005asynchronous, miller2016honey} to optimize block broadcasting during the broadcast stage.}

\begin{figure}[!tp]
    \centering
    \begin{subfigure}{0.49\linewidth}
    \centering
        \includegraphics[width=\linewidth]{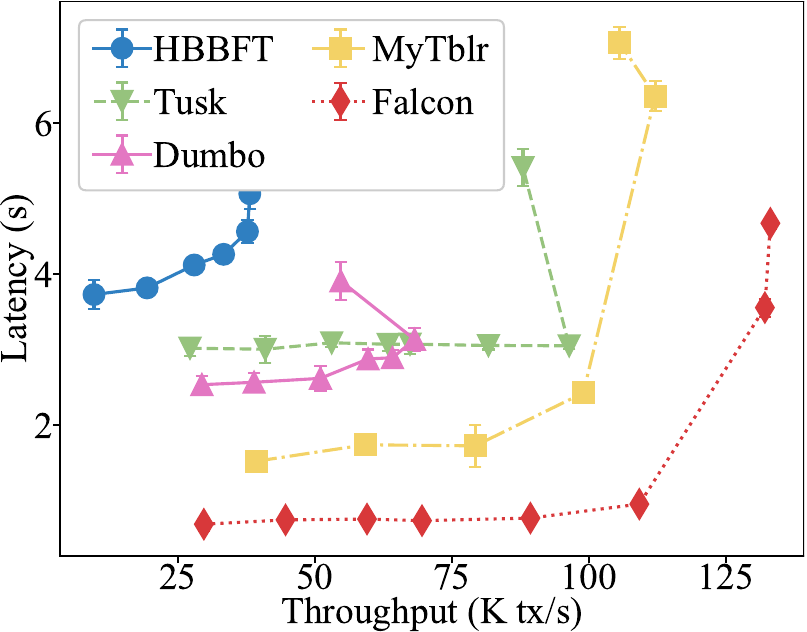}
        \vspace{-0.55cm}
        \caption{\rev{7 nodes}}
        \label{fig:perf-comp-fav-7}
    \end{subfigure}
    \begin{subfigure}{0.49\linewidth}
    \centering
        \includegraphics[width=\linewidth]{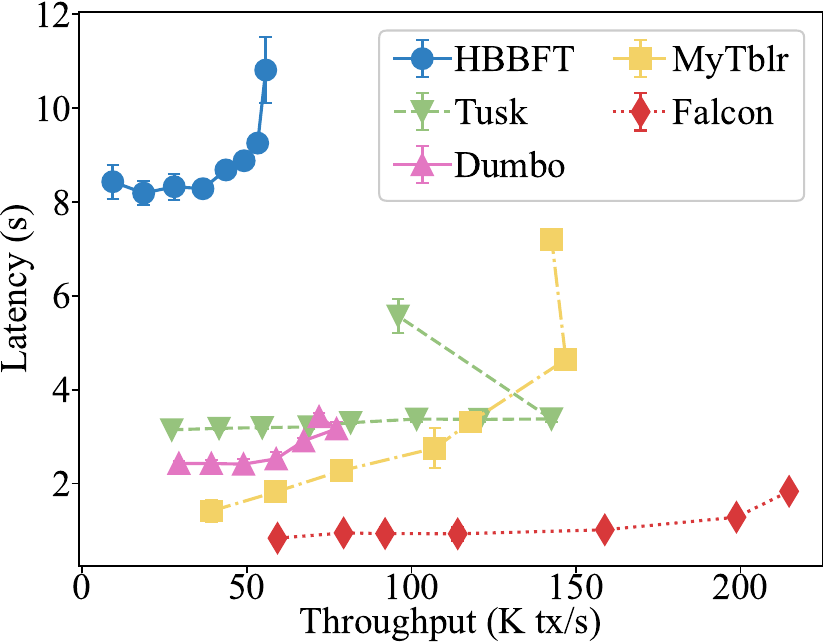}
        \vspace{-0.55cm}
        \caption{\rev{16 nodes}}
        \label{fig:perf-comp-fav-16}
    \end{subfigure}
    \vspace{-0.4cm}
    \caption{\rev{Latency \textit{v.s.} throughput in favorable situations}}
    \label{fig:perf-comp-fav}
\end{figure}
\begin{figure}[!tp]
    \centering
    \begin{subfigure}{0.49\linewidth}
    \centering
        \includegraphics[width=\linewidth]{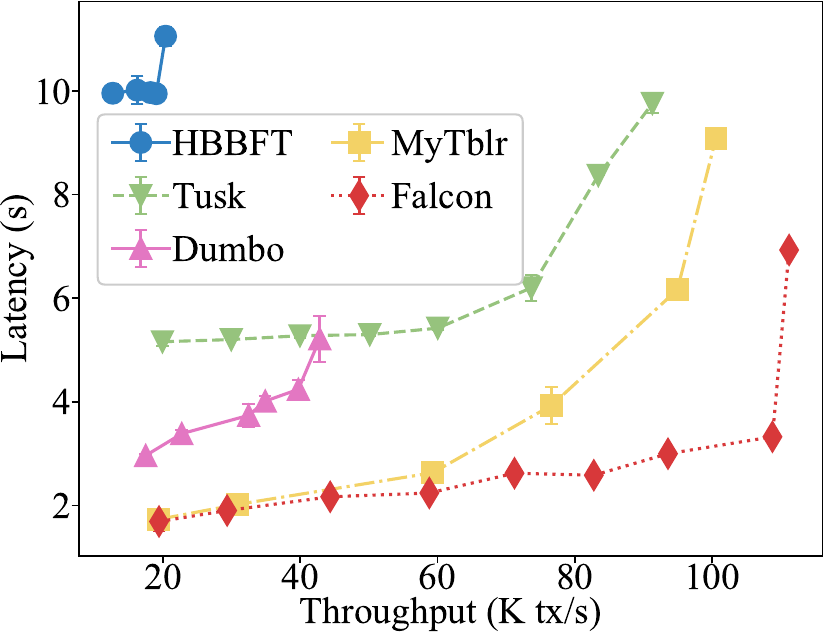}
        \vspace{-0.55cm}
        \caption{\rev{7 nodes}}
        \label{fig:perf-comp-unfav-7}
    \end{subfigure}
    \begin{subfigure}{0.49\linewidth}
    \centering
        \includegraphics[width=\linewidth]{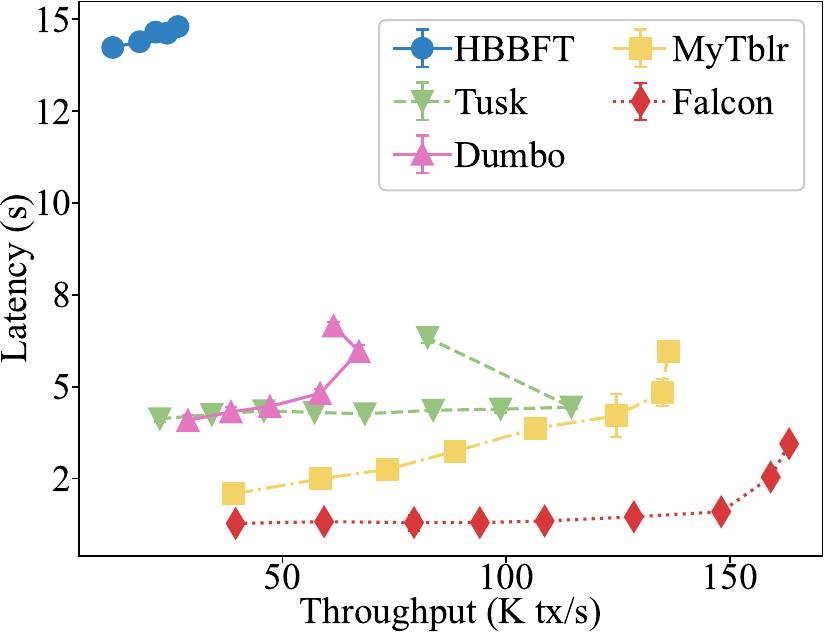}
        \vspace{-0.55cm}
        \caption{\rev{16 nodes}}
        \label{fig:perf-comp-unfav-16}
    \end{subfigure}
    \vspace{-0.4cm}
    \caption{\rev{Latency \textit{v.s.} throughput in unfavorable situations}}
    \label{fig:perf-comp-unfav}
\end{figure}

\subsection{Basic performance}
We employ two settings: one with 7 nodes and the other with 16 nodes. In each setting, we progressively increase the rate of client input transactions until the system reaches saturation. 

\subsubsection{Performance in favorable situations}
The trade-off between latency and throughput in favorable situations is shown in Figure~\ref{fig:perf-comp-fav}.
It is evident from the graph that \proto~exhibits lower baseline latency and higher peak throughput compared to other protocols. \rev{Specifically, when the system consists of 7 nodes, \proto's baseline latency is only 38.4\% of MyTblr, 24.2\% of Dumbo, 17.3\% of HBBFT, and 23.9\% of Tusk.} 
This advantage arises from two factors. First, \proto\ avoids running the agreement protocol in favorable situations, while Dumbo and HBBFT always need to execute it. 
\rev{As for Tusk, it relies on Narwhal to broadcast blocks and then commits them using a DAG-based agreement. Therefore, it also requires the execution of agreement protocols.}
Second, \proto's partial-sorting mechanism allows faster block committing, whereas HBBFT's integral-sorting requires all blocks to be decided before committing. MyTblr, with its timestamp-based sorting, requires blocks to wait until the timestamp reaches a specific threshold for committing.
Furthermore, regarding peak throughput, \proto~achieves a peak rate of 132.1K tx/s, \rev{surpassing MyTblr's 112.1K tx/s, Dumbo's 68.2K tx/s, HBBFT's 38.3K tx/s, and Tusk's 96.4K tx/s}. The high throughput of \proto~is attributed to its ability to commit more blocks for each ACSQ instance in favorable situations.

\subsubsection{Performance in unfavorable situations}
In unfavorable situations, we set the number of faulty nodes to 2 and 3 in the 7-node and 16-node settings, respectively. The experimental results are shown in Figure~\ref{fig:perf-comp-unfav}. As expected, the performance of all protocols declines when faulty nodes are present in the system compared to Figure~\ref{fig:perf-comp-fav}. \rev{However, \proto~demonstrates better performance than HBBFT, Dumbo, and Tusk. Specifically, in the 7-node setting, \proto's baseline latency is only 29.4\%, 70.7\%, 49.7\% of that of HBBFT, Dumbo, and Tusk, respectively.} This is due to the shortcut mechanism introduced in \proto's AABA protocol during the agreement stage, which allows AABA instances corresponding to faulty nodes to output more quickly. MyTblr exhibits a latency performance similar to \proto's because it also incorporates a fast path during the agreement stage, akin to the shortcut mechanism.

\begin{figure}[!tp]
    \centering
    \begin{subfigure}{0.49\linewidth}
    \centering
        \includegraphics[width=\linewidth]{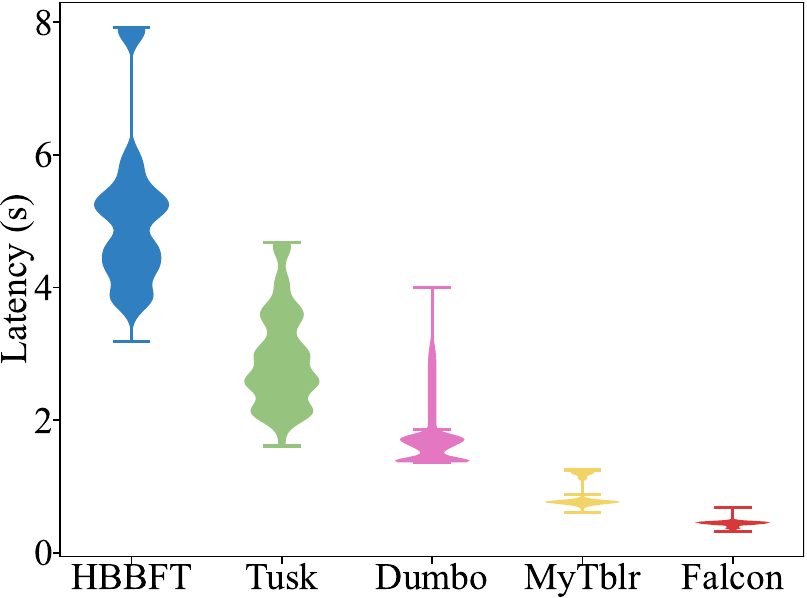}
        \vspace{-0.45cm}
        \caption{\rev{No faulty nodes}}
        \label{fig:lat-stab-fav}
    \end{subfigure}
    \begin{subfigure}{0.49\linewidth}
    \centering
        \includegraphics[width=\linewidth]{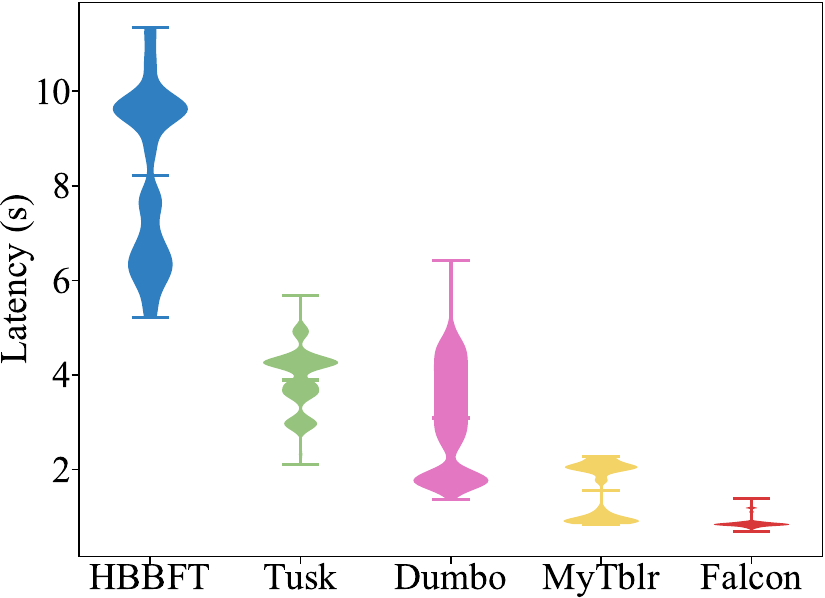}
        \vspace{-0.45cm}
        \caption{\rev{3 faulty nodes}}
        \label{fig:lat-stab-unfavor}
    \end{subfigure}
    \vspace{-0.2cm}
    \caption{\rev{Comparison of latency stability}}
    \vspace{-0.2cm}
    \label{fig:lat-stab}
\end{figure}

\begin{revpara}

\subsection{Latency stability}
We consider 16 nodes and examine both favorable and unfavorable situations. 
Experimental results are shown in Figure~\ref{fig:lat-stab}.

As shown in the figure, regardless of the situation, \proto~is able to continuously commit transactions due to the partial-sorting mechanism, ensuring a smaller latency range. In contrast, Dumbo, HBBFT, and Tusk only commit transactions intermittently, experiencing periods of no activity followed by large bursts of transactions, causing significant instability in latency. Additionally, performance comparisons across different scenarios indicate that the presence of faulty nodes worsens HBBFT’s performance, whereas \proto~consistently maintains stable latency.
Similarly, MyTblr also achieves relatively stable latency through its timestamp-based sorting mechanism;
however, this timestamp-based mechanism requires the timestamp to reach a specific threshold before block committing, leading to higher latency and reduced throughput.
\end{revpara}

\subsection{Latency decomposition}
We break down the latency to analyze the time consumed by each stage, namely the broadcast, agreement, and sorting. 
\rev{Note that while Tusk does not employ the ABA protocol, it achieves agreement through the leader's reference count mechanism. Therefore, we decompose the latency in Tusk as follows: broadcast time (block dissemination), agreement time (from the completion of block dissemination to when the leader meets the reference count condition), and sorting time (from the leader meeting the condition to the completion of block sorting).}

\begin{figure}[!tp]
    \centering
    \begin{subfigure}{0.49\linewidth}
    \centering
        \includegraphics[width=\linewidth]{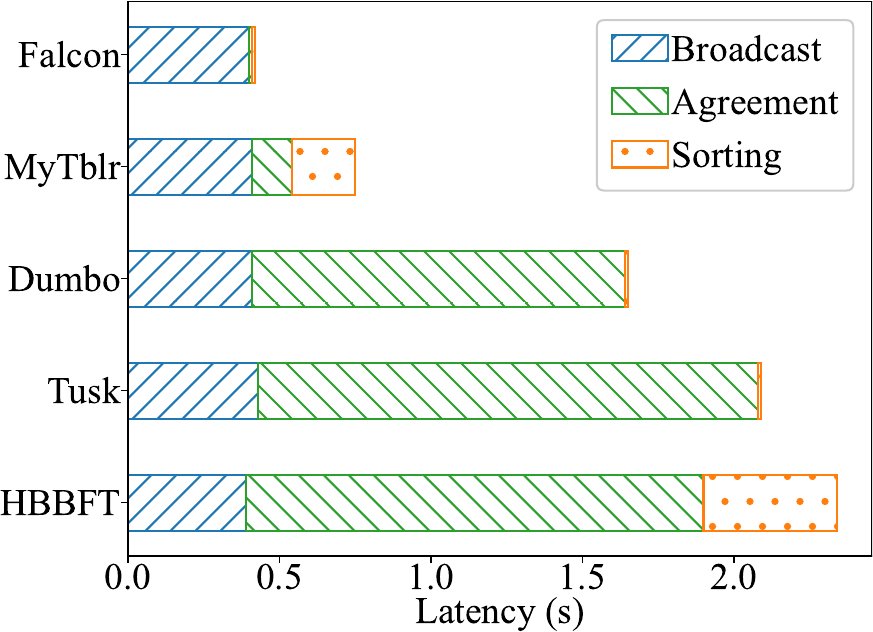}
        \vspace{-0.55cm}
        \caption{\rev{7 nodes}}
        \label{fig:7node-lat-decmp}
    \end{subfigure}
    \begin{subfigure}{0.49\linewidth}
    \centering
        \includegraphics[width=\linewidth]{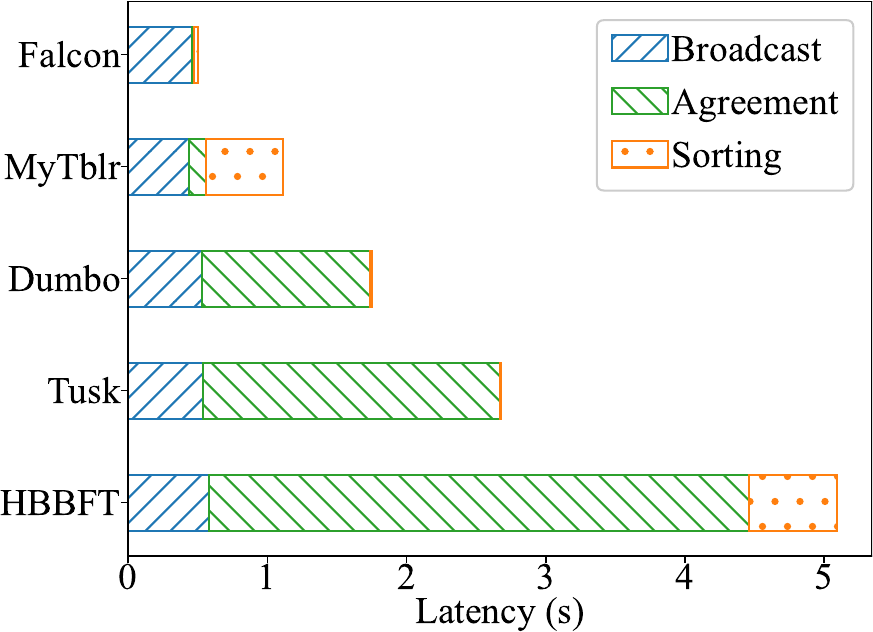}
        \vspace{-0.55cm}
        \caption{\rev{16 nodes}}
        \label{fig:16node-lat-decmp}
    \end{subfigure}
    \vspace{-0.4cm}
    \caption{\rev{Latency decomposition in favorable situations}}
    \label{fig:lat-decomp}
\end{figure}

The experimental results for favorable situations are shown in Figure~\ref{fig:lat-decomp}. It is evident that, regardless of the setup, the latency incurred by \proto~during the agreement stage is negligible. This validates \proto's design of using GBC for direct block committing, eliminating the agreement stage. 
\rev{In contrast, the agreement stage in the other four protocols takes a significant amount of time, particularly in HBBFT, Dumbo, and Tusk. In the 7-node system, the time spent in the agreement stage of HBBFT, Dumbo, and Tusk is 3.9 times, 2.7 times, and 3.8 times that of the broadcast stage, respectively.}
Moreover, \proto's sorting stage is also negligible due to its partial-sorting mechanism, which accelerates block committing by eliminating the need to wait for all blocks to be decided.


\begin{figure}[!tp]
    \centering
    \begin{subfigure}{0.49\linewidth}
    \centering
        \includegraphics[width=\linewidth]{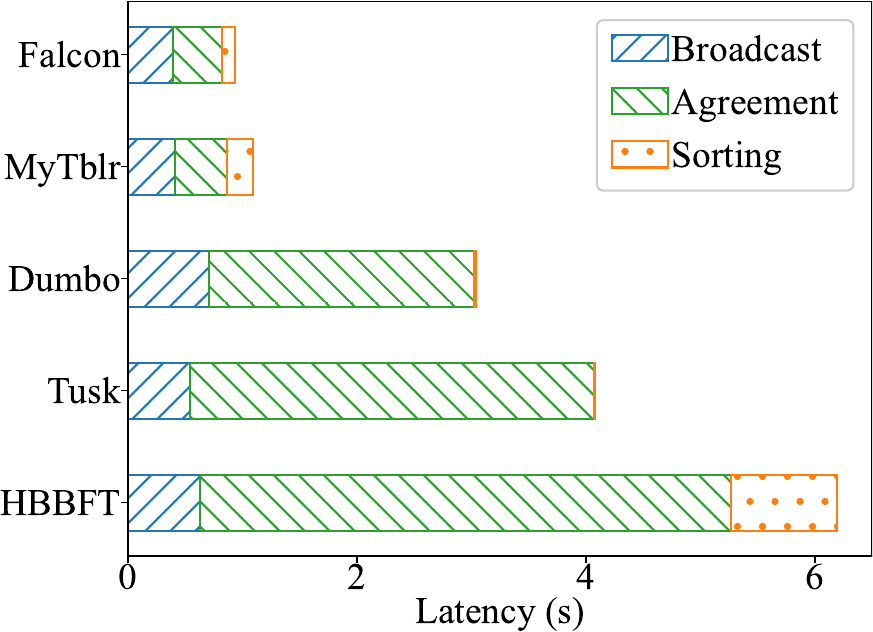}
        \vspace{-0.45cm}
        \caption{\rev{7 nodes}}
        \label{fig:7node-lat-decmp-unfav}
    \end{subfigure}
    \begin{subfigure}{0.49\linewidth}
    \centering
        \includegraphics[width=\linewidth]{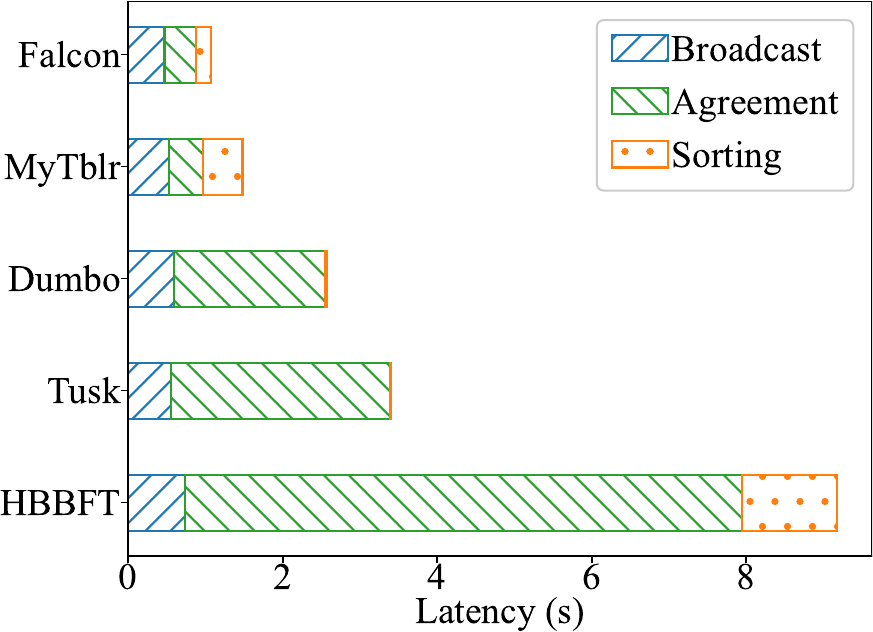}
        \vspace{-0.45cm}
        \caption{\rev{16 nodes}}
        \label{fig:16node-lat-decmp-unfav}
    \end{subfigure}
    \vspace{-0.4cm}
    \caption{\rev{Latency decomposition in unfavorable situations}}
    \label{fig:lat-decomp-unfav}
\end{figure}

The results for unfavorable situations are shown in Figure~\ref{fig:lat-decomp-unfav}. Due to the presence of faulty nodes, \proto~needs to run the agreement stage to commit blocks. However, with the shortcut mechanism introduced by AABA, \proto's agreement stage takes less time than others. \rev{Specifically, in the 16-node setting, \proto's agreement stage consumes only 80.1\%, 26.2\%, 6.1\%, and 14.4\% of the time required by MyTblr, Dumbo, HBBFT, and Tusk, respectively.}
As for the sorting stage, \proto's partial-sorting mechanism continues to reduce latency compared with MyTblr and HBBFT.

\rev{An interesting observation is that both Dumbo and Tusk exhibit almost no sorting time, regardless of whether the situations are favorable or unfavorable. This is because Dumbo employs the MVBA agreement protocol, where all blocks are decided simultaneously. 
Similarly, Tusk's reference count mechanism allows all blocks referenced by the leader to be simultaneously decided.}
The sorting process following simultaneous decisions is trivial and introduces minimal delay. However, the agreement process itself tends to incur relatively high latency.

\subsection{Scalability}

\begin{figure}[!tp]
    \centering
    \begin{subfigure}{0.48\linewidth}
    \centering
        \includegraphics[width=\linewidth]{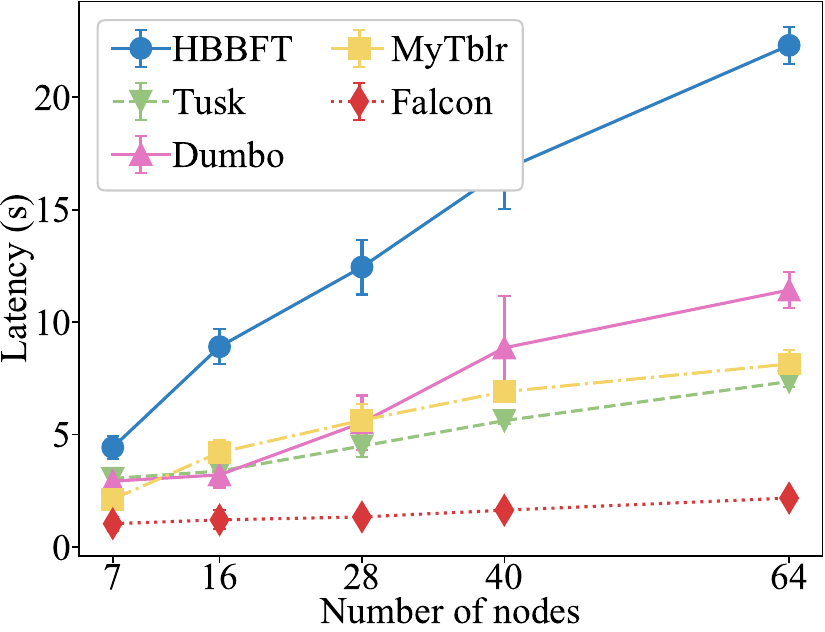}
        \vspace{-0.55cm}
        \caption{\rev{Latency comparison}}
        \label{fig:scale-latency}
    \end{subfigure}
    \begin{subfigure}{0.495\linewidth}
    \centering
        \includegraphics[width=\linewidth]{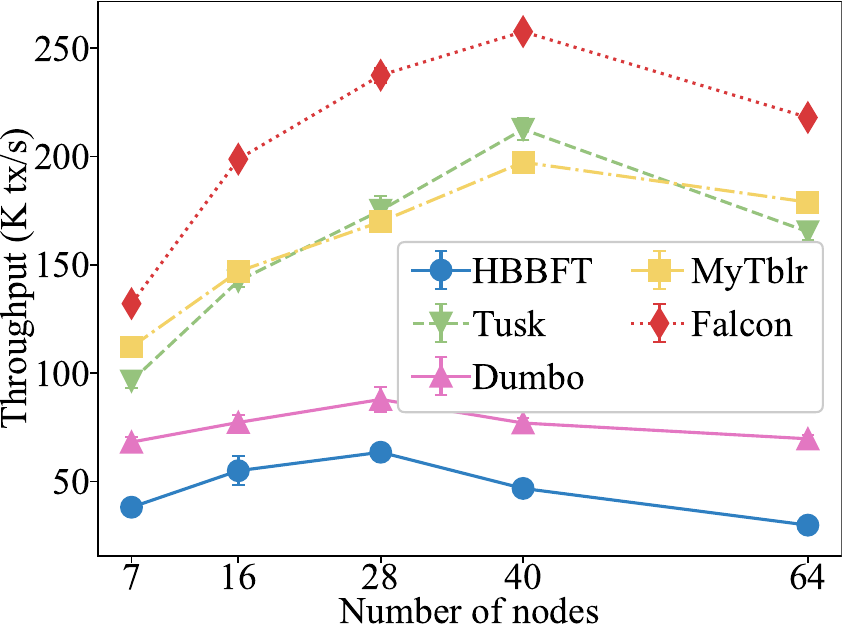}
        \vspace{-0.55cm}
        \caption{\rev{Throughput comparison}}
        \label{fig:scale-throughput}
    \end{subfigure}
    \vspace{-0.4cm}
    \caption{\rev{Performance comparison as node count increases}}
    \vspace{-0.5cm}
    \label{fig:scale}
\end{figure}

We evaluate the scalability of each protocol by increasing the number of nodes and analyzing the corresponding performance variations. Specifically, we focus on the optimistic scenario and, for each experiment, record the latency and throughput at the point of system saturation. 
The results are shown in Figure~\ref{fig:scale}.

As shown in Figure~\ref{fig:scale-latency}, while the latency of all protocols increases with node count, \proto~exhibits a significantly smaller increase compared to the others. For instance, when the number of nodes increases from 7 to 64, \proto's latency increases by only 2.1 times, \rev{whereas MyTblr, Dumbo, HBBFT, and Tusk experience increases of 3.8, 3.9, 5.1, and 2.4 times, respectively. Even with 64 nodes, \proto's latency remains at 2.2 seconds, which is notably lower than MyTblr's 8.2 seconds, Dumbo's 11.4 seconds, HBBFT's 22.3 seconds, and Tusk's 7.361 seconds.}

In terms of throughput, \proto~consistently outperforms others across different node scales. \rev{At 64 nodes, \proto\ achieves a throughput of 217.9K tx/s, surpassing MyTblr's 178.9K tx/s and Tusk's 165.3K tx/s, and significantly exceeding Dumbo's 69.7 K tx/s and HBBFT's 29.8K tx/s.}
Figure~\ref{fig:scale-throughput} shows that the throughput of all protocols initially increases and then decreases as the number of nodes grows. \rev{This pattern is linked to the mempool mechanism in \proto, MyTblr, Dumbo, and HBBFT, where each node generates and broadcasts payloads. In the early period, more nodes result in more payloads, boosting throughput. However, as the number of nodes increases further, the surge in consensus and payload messages leads to network congestion, which ultimately reduces throughput. 
For Tusk, a similar reasoning applies, as it depends on the Narwhal mempool.}
Despite this, \proto~consistently maintains higher throughput than the other protocols throughout all periods.

\section{Related Work}
In this section, we summarize the related work on asynchronous BFT protocols, while deferring the discussion of (partially) synchronous protocols to \autoref{sec:other-relat-work} due to space constraints.

Asynchronous consensus protocols include two key components: ABA~\cite{ben1983another, friedman2005simple, mostefaoui2014signature, abraham2022efficient} and MVBA~\cite{cachin2001secure, abraham2019asymptotically, lu2020dumbo, guo2022speeding}. ABA handles binary values, while MVBA handles arbitrary values. Combined with RBC or the GBC protocol proposed in this paper, ABA and MVBA can be used to implement an asynchronous BFT consensus, known as the BKR and CKPS paradigms, respectively.
As a representative of the BKR paradigm, HoneybadgerBFT was the first asynchronous BFT protocol to be practically deployed~\cite{miller2016honey}. 

\rev{Building on HoneybadgerBFT, Duan et al. introduced five versions of the BEAT protocol to optimize it~\cite{duan2018beat}. BEAT0 and BEAT2 optimize cryptographic protocols, which are orthogonal to our work and can be directly applied to our approach. BEAT3 and BEAT4 focus on storage system optimizations, which diverge from our goal of general SMR. BEAT1 suggests replacing the encoded RBC in HoneybadgerBFT with a non-encoded standard RBC. As noted in Section~\ref{sec:impl-set}, our block broadcasting implementation of Honeybadger has used a standard RBC rather than an encoded one.}
Liu et al. proposed a new agreement protocol, SuperMA, to improve the performance of the BKR paradigm~\cite{liu2023flexible}. However, even in optimistic situations, this protocol still requires running the ABA protocol. In contrast, \proto~avoids the need to run the ABA protocol in optimistic situations, resulting in better performance.

Some works have introduced an optimistic path into asynchronous protocols to enhance their performance in favorable conditions~\cite{ramasamy2005parsimonious, kursawe2005optimistic}. Depending on how the optimistic and pessimistic paths are structured, these approaches can be classified into the serial path paradigm~\cite{gelashvili2022jolteon, lu2022bolt} and the parallel path paradigm~\cite{blum2023abraxas, dai2023parbft}. They, however, can only commit a single block in optimistic situations, resulting in lower throughput. In contrast, \proto~can commit $n$ blocks in optimistic conditions, offering higher throughput.

Additionally, works like DAGRider~\cite{keidar2021all} have incorporated the topology of \textit{Directed Acyclic Graph} (DAG) into the design of consensus protocols. Building on DAGRider, protocols such as Tusk~\cite{danezis2022narwhal}, BullShark~\cite{spiegelman2022bullshark}, GradedDAG~\cite{gradeddag}, and Wahoo~\cite{dai2024wahoo} aim to optimize the latency of DAG-based BFT consensus. However, these protocols focus primarily on leader blocks and neglect non-leader blocks, which results in higher latency for non-leader blocks.

\section{Conclusion}
To address the key challenges of existing asynchronous BFT consensus, we present \proto, a protocol that offers both low and stable latency, as well as improved throughput.
Specifically, by introducing the GBC protocol, \proto~effectively bypasses the agreement stage in favorable situations, significantly reducing latency.
The AABA protocol ensures consistency between decisions made via GBC and those made during the agreement stage. Additionally, \proto's partial-sorting mechanism enhances latency stability by enabling continuous block committing. The integration of an agreement trigger further boosts throughput by allowing nodes to deliver and commit more blocks.
Experimental results demonstrate \proto's superiority over existing protocols, positioning it as a promising solution for advancing BFT consensus. 

\begin{acks}
 This work was supported by National Science and Technology Major Project 2022ZD0115301.
\end{acks}

\bibliographystyle{ACM-Reference-Format}
\bibliography{arxiv}


\begin{thebibliography}{58}


\ifx \showCODEN    \undefined \def \showCODEN     #1{\unskip}     \fi
\ifx \showDOI      \undefined \def \showDOI       #1{#1}\fi
\ifx \showISBNx    \undefined \def \showISBNx     #1{\unskip}     \fi
\ifx \showISBNxiii \undefined \def \showISBNxiii  #1{\unskip}     \fi
\ifx \showISSN     \undefined \def \showISSN      #1{\unskip}     \fi
\ifx \showLCCN     \undefined \def \showLCCN      #1{\unskip}     \fi
\ifx \shownote     \undefined \def \shownote      #1{#1}          \fi
\ifx \showarticletitle \undefined \def \showarticletitle #1{#1}   \fi
\ifx \showURL      \undefined \def \showURL       {\relax}        \fi
\providecommand\bibfield[2]{#2}
\providecommand\bibinfo[2]{#2}
\providecommand\natexlab[1]{#1}
\providecommand\showeprint[2][]{arXiv:#2}

\bibitem[\protect\citeauthoryear{Abraham and Asharov}{Abraham and
  Asharov}{2022}]%
        {abraham2022gradecast}
\bibfield{author}{\bibinfo{person}{Ittai Abraham} {and} \bibinfo{person}{Gilad
  Asharov}.} \bibinfo{year}{2022}\natexlab{}.
\newblock \showarticletitle{Gradecast in Synchrony and Reliable Broadcast in
  Asynchrony with Optimal Resilience, Efficiency, and Unconditional Security}.
  In \bibinfo{booktitle}{\emph{Proceedings of the 41st ACM Symposium on
  Principles of Distributed Computing}}. \bibinfo{pages}{392--398}.
\newblock


\bibitem[\protect\citeauthoryear{Abraham, Ben-David, and Yandamuri}{Abraham
  et~al\mbox{.}}{2022}]%
        {abraham2022efficient}
\bibfield{author}{\bibinfo{person}{Ittai Abraham}, \bibinfo{person}{Naama
  Ben-David}, {and} \bibinfo{person}{Sravya Yandamuri}.}
  \bibinfo{year}{2022}\natexlab{}.
\newblock \showarticletitle{Efficient and Adaptively Secure Asynchronous Binary
  Agreement via Binding Crusader Agreement}. In
  \bibinfo{booktitle}{\emph{Proceedings of the 41st ACM Symposium on Principles
  of Distributed Computing}}. ACM, \bibinfo{pages}{381--391}.
\newblock


\bibitem[\protect\citeauthoryear{Abraham, Malkhi, Nayak, Ren, and Yin}{Abraham
  et~al\mbox{.}}{2020}]%
        {abraham2020sync}
\bibfield{author}{\bibinfo{person}{Ittai Abraham}, \bibinfo{person}{Dahlia
  Malkhi}, \bibinfo{person}{Kartik Nayak}, \bibinfo{person}{Ling Ren}, {and}
  \bibinfo{person}{Maofan Yin}.} \bibinfo{year}{2020}\natexlab{}.
\newblock \showarticletitle{Sync Hotstuff: Simple and Practical Synchronous
  State Machine Replication}. In \bibinfo{booktitle}{\emph{Proceedings of the
  41st IEEE Symposium on Security and Privacy}}. IEEE,
  \bibinfo{pages}{106--118}.
\newblock


\bibitem[\protect\citeauthoryear{Abraham, Malkhi, and Spiegelman}{Abraham
  et~al\mbox{.}}{2019}]%
        {abraham2019asymptotically}
\bibfield{author}{\bibinfo{person}{Ittai Abraham}, \bibinfo{person}{Dahlia
  Malkhi}, {and} \bibinfo{person}{Alexander Spiegelman}.}
  \bibinfo{year}{2019}\natexlab{}.
\newblock \showarticletitle{Asymptotically Optimal Validated Asynchronous
  \text{Byzantine} Agreement}. In \bibinfo{booktitle}{\emph{Proceedings of the
  38th ACM Symposium on Principles of Distributed Computing}}.
  \bibinfo{pages}{337--346}.
\newblock


\bibitem[\protect\citeauthoryear{Amiri, Loo, Agrawal, and Abbadi}{Amiri
  et~al\mbox{.}}{2022}]%
        {amiri2022qanaat}
\bibfield{author}{\bibinfo{person}{Mohammad~Javad Amiri},
  \bibinfo{person}{Boon~Thau Loo}, \bibinfo{person}{Divyakant Agrawal}, {and}
  \bibinfo{person}{Amr~El Abbadi}.} \bibinfo{year}{2022}\natexlab{}.
\newblock \showarticletitle{Qanaat: A Scalable Multi-enterprise Permissioned
  Blockchain System with Confidentiality Guarantees}.
\newblock \bibinfo{journal}{\emph{Proceedings of the VLDB Endowment}}
  \bibinfo{volume}{15}, \bibinfo{number}{11} (\bibinfo{year}{2022}),
  \bibinfo{pages}{2839--2852}.
\newblock


\bibitem[\protect\citeauthoryear{Amoussou-Guenou, Del~Pozzo, Potop-Butucaru,
  and Tucci-Piergiovanni}{Amoussou-Guenou et~al\mbox{.}}{2019}]%
        {amoussou2019dissecting}
\bibfield{author}{\bibinfo{person}{Yackolley Amoussou-Guenou},
  \bibinfo{person}{Antonella Del~Pozzo}, \bibinfo{person}{Maria
  Potop-Butucaru}, {and} \bibinfo{person}{Sara Tucci-Piergiovanni}.}
  \bibinfo{year}{2019}\natexlab{}.
\newblock \showarticletitle{Dissecting \text{Tendermint}}. In
  \bibinfo{booktitle}{\emph{Proceedings of the 7th International Conference on
  Networked Systems}}. Springer, \bibinfo{pages}{166--182}.
\newblock


\bibitem[\protect\citeauthoryear{Arun and Ravindran}{Arun and
  Ravindran}{2022}]%
        {arun2022scalable}
\bibfield{author}{\bibinfo{person}{Balaji Arun} {and} \bibinfo{person}{Binoy
  Ravindran}.} \bibinfo{year}{2022}\natexlab{}.
\newblock \showarticletitle{Scalable Byzantine Fault Tolerance via Partial
  Decentralization}.
\newblock \bibinfo{journal}{\emph{Proceedings of the VLDB Endowment}}
  \bibinfo{volume}{15}, \bibinfo{number}{9} (\bibinfo{year}{2022}),
  \bibinfo{pages}{1739–1752}.
\newblock


\bibitem[\protect\citeauthoryear{Ben-Or}{Ben-Or}{1983}]%
        {ben1983another}
\bibfield{author}{\bibinfo{person}{Michael Ben-Or}.}
  \bibinfo{year}{1983}\natexlab{}.
\newblock \showarticletitle{Another Advantage of Free Choice: Completely
  Asynchronous Agreement Protocols}. In \bibinfo{booktitle}{\emph{Proceedings
  of the 2nd Annual ACM Symposium on Principles of Distributed Computing}}.
  ACM, \bibinfo{pages}{27--30}.
\newblock


\bibitem[\protect\citeauthoryear{Ben-Or, Kelmer, and Rabin}{Ben-Or
  et~al\mbox{.}}{1994}]%
        {ben1994asynchronous}
\bibfield{author}{\bibinfo{person}{Michael Ben-Or}, \bibinfo{person}{Boaz
  Kelmer}, {and} \bibinfo{person}{Tal Rabin}.} \bibinfo{year}{1994}\natexlab{}.
\newblock \showarticletitle{Asynchronous Secure Computations with Optimal
  Resilience}. In \bibinfo{booktitle}{\emph{Proceedings of the 13th Annual ACM
  Symposium on Principles of Distributed Computing}}.
  \bibinfo{pages}{183--192}.
\newblock


\bibitem[\protect\citeauthoryear{Blum, Katz, Loss, Nayak, and
  Ochsenreither}{Blum et~al\mbox{.}}{2023}]%
        {blum2023abraxas}
\bibfield{author}{\bibinfo{person}{Erica Blum}, \bibinfo{person}{Jonathan
  Katz}, \bibinfo{person}{Julian Loss}, \bibinfo{person}{Kartik Nayak}, {and}
  \bibinfo{person}{Simon Ochsenreither}.} \bibinfo{year}{2023}\natexlab{}.
\newblock \showarticletitle{Abraxas: Throughput-Efficient Hybrid Asynchronous
  Consensus}. In \bibinfo{booktitle}{\emph{Proceedings of the 30th ACM
  Conference on Computer and Communications Security}}.
  \bibinfo{pages}{519--533}.
\newblock


\bibitem[\protect\citeauthoryear{Bracha}{Bracha}{1987}]%
        {bracha1987asynchronous}
\bibfield{author}{\bibinfo{person}{Gabriel Bracha}.}
  \bibinfo{year}{1987}\natexlab{}.
\newblock \showarticletitle{Asynchronous Byzantine Agreement Protocols}.
\newblock \bibinfo{journal}{\emph{Information and Computation}}
  \bibinfo{volume}{75}, \bibinfo{number}{2} (\bibinfo{year}{1987}),
  \bibinfo{pages}{130--143}.
\newblock


\bibitem[\protect\citeauthoryear{Buchnik and Friedman}{Buchnik and
  Friedman}{2020}]%
        {buchnik2020fireledger}
\bibfield{author}{\bibinfo{person}{Yehonatan Buchnik} {and}
  \bibinfo{person}{Roy Friedman}.} \bibinfo{year}{2020}\natexlab{}.
\newblock \showarticletitle{FireLedger: A High Throughput Blockchain Consensus
  Protocol}.
\newblock \bibinfo{journal}{\emph{Proceedings of the VLDB Endowment}}
  \bibinfo{volume}{13}, \bibinfo{number}{9} (\bibinfo{year}{2020}),
  \bibinfo{pages}{1525--1539}.
\newblock


\bibitem[\protect\citeauthoryear{Cachin, Kursawe, Petzold, and Shoup}{Cachin
  et~al\mbox{.}}{2001}]%
        {cachin2001secure}
\bibfield{author}{\bibinfo{person}{Christian Cachin}, \bibinfo{person}{Klaus
  Kursawe}, \bibinfo{person}{Frank Petzold}, {and} \bibinfo{person}{Victor
  Shoup}.} \bibinfo{year}{2001}\natexlab{}.
\newblock \showarticletitle{Secure and Efficient Asynchronous Broadcast
  Protocols}. In \bibinfo{booktitle}{\emph{Proceedings of the 21st Annual
  International Cryptology Conference}}. Springer, \bibinfo{pages}{524--541}.
\newblock


\bibitem[\protect\citeauthoryear{Cachin and Tessaro}{Cachin and
  Tessaro}{2005}]%
        {cachin2005asynchronous}
\bibfield{author}{\bibinfo{person}{Christian Cachin} {and}
  \bibinfo{person}{Stefano Tessaro}.} \bibinfo{year}{2005}\natexlab{}.
\newblock \showarticletitle{Asynchronous Verifiable Information Dispersal}. In
  \bibinfo{booktitle}{\emph{Proceedings of the 24th IEEE Symposium on Reliable
  Distributed Systems}}. IEEE, \bibinfo{pages}{191--201}.
\newblock


\bibitem[\protect\citeauthoryear{Castro and Liskov}{Castro and Liskov}{1999}]%
        {castro1999practical}
\bibfield{author}{\bibinfo{person}{Miguel Castro} {and}
  \bibinfo{person}{Barbara Liskov}.} \bibinfo{year}{1999}\natexlab{}.
\newblock \showarticletitle{Practical \text{Byzantine} Fault Tolerance}. In
  \bibinfo{booktitle}{\emph{Proceedings of the 3rd USENIX Symposium on
  Operating Systems Design and Implementation}}. USENIX,
  \bibinfo{pages}{173--186}.
\newblock


\bibitem[\protect\citeauthoryear{Chan, Pass, and Shi}{Chan
  et~al\mbox{.}}{2018}]%
        {chan2018pili}
\bibfield{author}{\bibinfo{person}{TH~Hubert Chan}, \bibinfo{person}{Rafael
  Pass}, {and} \bibinfo{person}{Elaine Shi}.} \bibinfo{year}{2018}\natexlab{}.
\newblock \showarticletitle{Pili: An Extremely Simple Synchronous Blockchain}.
\newblock \bibinfo{journal}{\emph{Cryptology ePrint Archive}}
  (\bibinfo{year}{2018}).
\newblock


\bibitem[\protect\citeauthoryear{Dai, Huang, Xiao, Zhang, Xie, and Jin}{Dai
  et~al\mbox{.}}{2022}]%
        {dai2022trebiz}
\bibfield{author}{\bibinfo{person}{Xiaohai Dai}, \bibinfo{person}{Liping
  Huang}, \bibinfo{person}{Jiang Xiao}, \bibinfo{person}{Zhaonan Zhang},
  \bibinfo{person}{Xia Xie}, {and} \bibinfo{person}{Hai Jin}.}
  \bibinfo{year}{2022}\natexlab{}.
\newblock \showarticletitle{Trebiz: \text{Byzantine} Fault Tolerance with
  \text{Byzantine} Merchants}. In \bibinfo{booktitle}{\emph{Proceedings of the
  38th Annual Computer Security Applications Conference}}. ACSA,
  \bibinfo{pages}{923--935}.
\newblock


\bibitem[\protect\citeauthoryear{Dai, Zhang, Jin, and Ren}{Dai
  et~al\mbox{.}}{2023a}]%
        {dai2023parbft}
\bibfield{author}{\bibinfo{person}{Xiaohai Dai}, \bibinfo{person}{Bolin Zhang},
  \bibinfo{person}{Hai Jin}, {and} \bibinfo{person}{Ling Ren}.}
  \bibinfo{year}{2023}\natexlab{a}.
\newblock \showarticletitle{\text{ParBFT}: Faster Asynchronous \text{BFT}
  Consensus with a Parallel Optimistic Path}. In
  \bibinfo{booktitle}{\emph{Proceedings of the 30th ACM Conference on Computer
  and Communications Security}}. \bibinfo{pages}{504--518}.
\newblock


\bibitem[\protect\citeauthoryear{Dai, Zhang, Guo, Ding, Xiao, Xie, Hao, and
  Jin}{Dai et~al\mbox{.}}{2024}]%
        {dai2024wahoo}
\bibfield{author}{\bibinfo{person}{Xiaohai Dai}, \bibinfo{person}{Zhaonan
  Zhang}, \bibinfo{person}{Zhengxuan Guo}, \bibinfo{person}{Chaozheng Ding},
  \bibinfo{person}{Jiang Xiao}, \bibinfo{person}{Xia Xie}, \bibinfo{person}{Rui
  Hao}, {and} \bibinfo{person}{Hai Jin}.} \bibinfo{year}{2024}\natexlab{}.
\newblock \showarticletitle{Wahoo: A DAG-Based BFT Consensus With Low Latency
  and Low Communication Overhead}.
\newblock \bibinfo{journal}{\emph{IEEE Transactions on Information Forensics
  and Security}}  \bibinfo{volume}{19} (\bibinfo{year}{2024}),
  \bibinfo{pages}{7508--7522}.
\newblock


\bibitem[\protect\citeauthoryear{Dai, Zhang, Xiao, Yue, Xie, and Jin}{Dai
  et~al\mbox{.}}{2023b}]%
        {gradeddag}
\bibfield{author}{\bibinfo{person}{Xiaohai Dai}, \bibinfo{person}{Zhaonan
  Zhang}, \bibinfo{person}{Jiang Xiao}, \bibinfo{person}{Jingtao Yue},
  \bibinfo{person}{Xia Xie}, {and} \bibinfo{person}{Hai Jin}.}
  \bibinfo{year}{2023}\natexlab{b}.
\newblock \showarticletitle{\text{GradedDAG}: An Asynchronous \text{DAG}-based
  \text{BFT} Consensus with Lower Latency}. In
  \bibinfo{booktitle}{\emph{Proceedings of the 42nd International Symposium on
  Reliable Distributed Systems}}. \bibinfo{pages}{107--117}.
\newblock


\bibitem[\protect\citeauthoryear{Danezis, Kokoris-Kogias, Sonnino, and
  Spiegelman}{Danezis et~al\mbox{.}}{2022}]%
        {danezis2022narwhal}
\bibfield{author}{\bibinfo{person}{George Danezis}, \bibinfo{person}{Lefteris
  Kokoris-Kogias}, \bibinfo{person}{Alberto Sonnino}, {and}
  \bibinfo{person}{Alexander Spiegelman}.} \bibinfo{year}{2022}\natexlab{}.
\newblock \showarticletitle{Narwhal and \text{Tusk}: A \text{DAG}-based Mempool
  and Efficient \text{BFT} consensus}. In \bibinfo{booktitle}{\emph{Proceedings
  of the 17th European Conference on Computer Systems}}. ACM,
  \bibinfo{pages}{34--50}.
\newblock


\bibitem[\protect\citeauthoryear{Duan, Reiter, and Zhang}{Duan
  et~al\mbox{.}}{2018}]%
        {duan2018beat}
\bibfield{author}{\bibinfo{person}{Sisi Duan}, \bibinfo{person}{Michael~K
  Reiter}, {and} \bibinfo{person}{Haibin Zhang}.}
  \bibinfo{year}{2018}\natexlab{}.
\newblock \showarticletitle{BEAT: Asynchronous BFT Made Practical}. In
  \bibinfo{booktitle}{\emph{Proceedings of the 25th ACM SIGSAC Conference on
  Computer and Communications Security}}. \bibinfo{pages}{2028--2041}.
\newblock


\bibitem[\protect\citeauthoryear{Dwork, Lynch, and Stockmeyer}{Dwork
  et~al\mbox{.}}{1988}]%
        {dwork1988consensus}
\bibfield{author}{\bibinfo{person}{Cynthia Dwork}, \bibinfo{person}{Nancy
  Lynch}, {and} \bibinfo{person}{Larry Stockmeyer}.}
  \bibinfo{year}{1988}\natexlab{}.
\newblock \showarticletitle{Consensus in the Presence of Partial Synchrony}.
\newblock \bibinfo{journal}{\emph{J. ACM}} \bibinfo{volume}{35},
  \bibinfo{number}{2} (\bibinfo{year}{1988}), \bibinfo{pages}{288--323}.
\newblock


\bibitem[\protect\citeauthoryear{Friedman, Mostefaoui, and Raynal}{Friedman
  et~al\mbox{.}}{2005}]%
        {friedman2005simple}
\bibfield{author}{\bibinfo{person}{Roy Friedman}, \bibinfo{person}{Achour
  Mostefaoui}, {and} \bibinfo{person}{Michel Raynal}.}
  \bibinfo{year}{2005}\natexlab{}.
\newblock \showarticletitle{Simple and Efficient Oracle-based Consensus
  Protocols for Asynchronous Byzantine Systems}.
\newblock \bibinfo{journal}{\emph{IEEE Transactions on Dependable and Secure
  Computing}} \bibinfo{volume}{2}, \bibinfo{number}{1} (\bibinfo{year}{2005}),
  \bibinfo{pages}{46--56}.
\newblock


\bibitem[\protect\citeauthoryear{Gai, Niu, Beschastnikh, Feng, and Wang}{Gai
  et~al\mbox{.}}{2023}]%
        {gai2023scaling}
\bibfield{author}{\bibinfo{person}{Fangyu Gai}, \bibinfo{person}{Jianyu Niu},
  \bibinfo{person}{Ivan Beschastnikh}, \bibinfo{person}{Chen Feng}, {and}
  \bibinfo{person}{Sheng Wang}.} \bibinfo{year}{2023}\natexlab{}.
\newblock \showarticletitle{Scaling blockchain consensus via a robust shared
  mempool}. In \bibinfo{booktitle}{\emph{2023 IEEE 39th International
  Conference on Data Engineering (ICDE)}}. IEEE, \bibinfo{pages}{530--543}.
\newblock


\bibitem[\protect\citeauthoryear{Gelashvili, Kokoris-Kogias, Sonnino,
  Spiegelman, and Xiang}{Gelashvili et~al\mbox{.}}{2022}]%
        {gelashvili2022jolteon}
\bibfield{author}{\bibinfo{person}{Rati Gelashvili}, \bibinfo{person}{Lefteris
  Kokoris-Kogias}, \bibinfo{person}{Alberto Sonnino},
  \bibinfo{person}{Alexander Spiegelman}, {and} \bibinfo{person}{Zhuolun
  Xiang}.} \bibinfo{year}{2022}\natexlab{}.
\newblock \showarticletitle{Jolteon and \text{Ditto}: Network-adaptive
  Efficient Consensus with Asynchronous Fallback}. In
  \bibinfo{booktitle}{\emph{Proceedings of the 26th International Conference on
  Financial Cryptography and Data Security}}. Springer,
  \bibinfo{pages}{296--315}.
\newblock


\bibitem[\protect\citeauthoryear{Gelashvili, Spiegelman, Xiang, Danezis, Li,
  Malkhi, Xia, and Zhou}{Gelashvili et~al\mbox{.}}{2023}]%
        {gelashvili2023block}
\bibfield{author}{\bibinfo{person}{Rati Gelashvili}, \bibinfo{person}{Alexander
  Spiegelman}, \bibinfo{person}{Zhuolun Xiang}, \bibinfo{person}{George
  Danezis}, \bibinfo{person}{Zekun Li}, \bibinfo{person}{Dahlia Malkhi},
  \bibinfo{person}{Yu Xia}, {and} \bibinfo{person}{Runtian Zhou}.}
  \bibinfo{year}{2023}\natexlab{}.
\newblock \showarticletitle{Block-stm: Scaling Blockchain Execution by Turning
  Ordering Curse to a Performance Blessing}. In
  \bibinfo{booktitle}{\emph{Proceedings of the 28th ACM SIGPLAN Annual
  Symposium on Principles and Practice of Parallel Programming}}.
  \bibinfo{pages}{232--244}.
\newblock


\bibitem[\protect\citeauthoryear{Gueta, Abraham, Grossman, Malkhi, Pinkas,
  Reiter, Seredinschi, Tamir, and Tomescu}{Gueta et~al\mbox{.}}{2019}]%
        {gueta2019sbft}
\bibfield{author}{\bibinfo{person}{Guy~Golan Gueta}, \bibinfo{person}{Ittai
  Abraham}, \bibinfo{person}{Shelly Grossman}, \bibinfo{person}{Dahlia Malkhi},
  \bibinfo{person}{Benny Pinkas}, \bibinfo{person}{Michael Reiter},
  \bibinfo{person}{Dragos-Adrian Seredinschi}, \bibinfo{person}{Orr Tamir},
  {and} \bibinfo{person}{Alin Tomescu}.} \bibinfo{year}{2019}\natexlab{}.
\newblock \showarticletitle{\text{SBFT}: A Scalable and Decentralized Trust
  Infrastructure}. In \bibinfo{booktitle}{\emph{Proceedings of the 49th Annual
  IEEE/IFIP International Conference on Dependable Systems and Networks}}.
  IEEE, \bibinfo{pages}{568--580}.
\newblock


\bibitem[\protect\citeauthoryear{Guo, Lu, Lu, Tang, Xu, and Zhang}{Guo
  et~al\mbox{.}}{2022}]%
        {guo2022speeding}
\bibfield{author}{\bibinfo{person}{Bingyong Guo}, \bibinfo{person}{Yuan Lu},
  \bibinfo{person}{Zhenliang Lu}, \bibinfo{person}{Qiang Tang},
  \bibinfo{person}{Jing Xu}, {and} \bibinfo{person}{Zhenfeng Zhang}.}
  \bibinfo{year}{2022}\natexlab{}.
\newblock \showarticletitle{Speeding Dumbo: Pushing Asynchronous BFT Closer to
  Practice}.
\newblock \bibinfo{journal}{\emph{Cryptology ePrint Archive}}
  (\bibinfo{year}{2022}).
\newblock


\bibitem[\protect\citeauthoryear{Guo, Lu, Tang, Xu, and Zhang}{Guo
  et~al\mbox{.}}{2020}]%
        {guo2020dumbo}
\bibfield{author}{\bibinfo{person}{Bingyong Guo}, \bibinfo{person}{Zhenliang
  Lu}, \bibinfo{person}{Qiang Tang}, \bibinfo{person}{Jing Xu}, {and}
  \bibinfo{person}{Zhenfeng Zhang}.} \bibinfo{year}{2020}\natexlab{}.
\newblock \showarticletitle{Dumbo: Faster Asynchronous BFT Protocols}. In
  \bibinfo{booktitle}{\emph{Proceedings of the 27th ACM SIGSAC Conference on
  Computer and Communications Security}}. \bibinfo{pages}{803--818}.
\newblock


\bibitem[\protect\citeauthoryear{Kang, Rahnama, Hellings, and Sadoghi}{Kang
  et~al\mbox{.}}{2024}]%
        {kang2024spotless}
\bibfield{author}{\bibinfo{person}{Dakai Kang}, \bibinfo{person}{Sajjad
  Rahnama}, \bibinfo{person}{Jelle Hellings}, {and} \bibinfo{person}{Mohammad
  Sadoghi}.} \bibinfo{year}{2024}\natexlab{}.
\newblock \showarticletitle{Spotless: Concurrent rotational consensus made
  practical through rapid view synchronization}. In
  \bibinfo{booktitle}{\emph{2024 IEEE 40th International Conference on Data
  Engineering (ICDE)}}. IEEE, \bibinfo{pages}{1916--1929}.
\newblock


\bibitem[\protect\citeauthoryear{Keidar, Kokoris-Kogias, Naor, and
  Spiegelman}{Keidar et~al\mbox{.}}{2021}]%
        {keidar2021all}
\bibfield{author}{\bibinfo{person}{Idit Keidar}, \bibinfo{person}{Eleftherios
  Kokoris-Kogias}, \bibinfo{person}{Oded Naor}, {and}
  \bibinfo{person}{Alexander Spiegelman}.} \bibinfo{year}{2021}\natexlab{}.
\newblock \showarticletitle{All You Need is \text{DAG}}. In
  \bibinfo{booktitle}{\emph{Proceedings of the 40th ACM Symposium on Principles
  of Distributed Computing}}. ACM, \bibinfo{pages}{165--175}.
\newblock


\bibitem[\protect\citeauthoryear{Kihlstrom, Moser, and Melliar-Smith}{Kihlstrom
  et~al\mbox{.}}{1998}]%
        {kihlstrom1998securering}
\bibfield{author}{\bibinfo{person}{Kim~Potter Kihlstrom},
  \bibinfo{person}{Louise~E Moser}, {and} \bibinfo{person}{P~Michael
  Melliar-Smith}.} \bibinfo{year}{1998}\natexlab{}.
\newblock \showarticletitle{The Secure Ring Protocols for Securing Group
  Communication}. In \bibinfo{booktitle}{\emph{Proceedings of the 31st Hawaii
  International Conference on System Sciences}}, Vol.~\bibinfo{volume}{3}.
  IEEE, \bibinfo{pages}{317--326}.
\newblock


\bibitem[\protect\citeauthoryear{Kotla, Alvisi, Dahlin, Clement, and
  Wong}{Kotla et~al\mbox{.}}{2007}]%
        {kotla2007zyzzyva}
\bibfield{author}{\bibinfo{person}{Ramakrishna Kotla}, \bibinfo{person}{Lorenzo
  Alvisi}, \bibinfo{person}{Mike Dahlin}, \bibinfo{person}{Allen Clement},
  {and} \bibinfo{person}{Edmund Wong}.} \bibinfo{year}{2007}\natexlab{}.
\newblock \showarticletitle{Zyzzyva: Speculative \text{Byzantine} Fault
  Tolerance}. In \bibinfo{booktitle}{\emph{Proceedings of the 13rd ACM SIGOPS
  Symposium on Operating Systems Principles}}. ACM, \bibinfo{pages}{45--58}.
\newblock


\bibitem[\protect\citeauthoryear{Kursawe and Shoup}{Kursawe and Shoup}{2005}]%
        {kursawe2005optimistic}
\bibfield{author}{\bibinfo{person}{Klaus Kursawe} {and} \bibinfo{person}{Victor
  Shoup}.} \bibinfo{year}{2005}\natexlab{}.
\newblock \showarticletitle{Optimistic Asynchronous Atomic Broadcast}. In
  \bibinfo{booktitle}{\emph{Proceedings of the 32nd International Colloquium on
  Automata, Languages, and Programming}}. Springer, \bibinfo{pages}{204--215}.
\newblock


\bibitem[\protect\citeauthoryear{Liu, Li, Karame, and Asokan}{Liu
  et~al\mbox{.}}{2018}]%
        {liu2018scalable}
\bibfield{author}{\bibinfo{person}{Jian Liu}, \bibinfo{person}{Wenting Li},
  \bibinfo{person}{Ghassan~O Karame}, {and} \bibinfo{person}{N Asokan}.}
  \bibinfo{year}{2018}\natexlab{}.
\newblock \showarticletitle{Scalable \text{Byzantine} Consensus via
  Hardware-assisted Secret Sharing}.
\newblock \bibinfo{journal}{\emph{IEEE Trans. Comput.}} \bibinfo{volume}{68},
  \bibinfo{number}{1} (\bibinfo{year}{2018}), \bibinfo{pages}{139--151}.
\newblock


\bibitem[\protect\citeauthoryear{Liu, Xu, Shan, Yan, Xu, Wang, Fan, Deng, Yan,
  and Zhang}{Liu et~al\mbox{.}}{2023}]%
        {liu2023flexible}
\bibfield{author}{\bibinfo{person}{Shengyun Liu}, \bibinfo{person}{Wenbo Xu},
  \bibinfo{person}{Chen Shan}, \bibinfo{person}{Xiaofeng Yan},
  \bibinfo{person}{Tianjing Xu}, \bibinfo{person}{Bo Wang},
  \bibinfo{person}{Lei Fan}, \bibinfo{person}{Fuxi Deng}, \bibinfo{person}{Ying
  Yan}, {and} \bibinfo{person}{Hui Zhang}.} \bibinfo{year}{2023}\natexlab{}.
\newblock \showarticletitle{Flexible Advancement in Asynchronous BFT
  Consensus}. In \bibinfo{booktitle}{\emph{Proceedings of the 29th Symposium on
  Operating Systems Principles}}. \bibinfo{pages}{264--280}.
\newblock


\bibitem[\protect\citeauthoryear{Lu, Lu, and Tang}{Lu et~al\mbox{.}}{2022}]%
        {lu2022bolt}
\bibfield{author}{\bibinfo{person}{Yuan Lu}, \bibinfo{person}{Zhenliang Lu},
  {and} \bibinfo{person}{Qiang Tang}.} \bibinfo{year}{2022}\natexlab{}.
\newblock \showarticletitle{\text{Bolt-Dumbo} Transformer: Asynchronous
  Consensus as Fast as the Pipelined \text{BFT}}. In
  \bibinfo{booktitle}{\emph{Proceedings of the 29th ACM Conference on Computer
  and Communications Security}}. \bibinfo{pages}{2159--2173}.
\newblock


\bibitem[\protect\citeauthoryear{Lu, Lu, Tang, and Wang}{Lu
  et~al\mbox{.}}{2020}]%
        {lu2020dumbo}
\bibfield{author}{\bibinfo{person}{Yuan Lu}, \bibinfo{person}{Zhenliang Lu},
  \bibinfo{person}{Qiang Tang}, {and} \bibinfo{person}{Guiling Wang}.}
  \bibinfo{year}{2020}\natexlab{}.
\newblock \showarticletitle{\text{Dumbo-MVBA}: Optimal Multi-valued Validated
  Asynchronous \text{Byzantine} Agreement, Revisited}. In
  \bibinfo{booktitle}{\emph{Proceedings of the 39th Symposium on Principles of
  Distributed Computing}}. \bibinfo{pages}{129--138}.
\newblock


\bibitem[\protect\citeauthoryear{Malkhi, Stathakopoulou, and Yin}{Malkhi
  et~al\mbox{.}}{2024}]%
        {malkhi2024bbca}
\bibfield{author}{\bibinfo{person}{Dahlia Malkhi}, \bibinfo{person}{Chrysoula
  Stathakopoulou}, {and} \bibinfo{person}{Maofan Yin}.}
  \bibinfo{year}{2024}\natexlab{}.
\newblock \showarticletitle{BBCA-CHAIN: Low latency, High Throughput BFT
  Consensus on a DAG}. In \bibinfo{booktitle}{\emph{Proceedings of the 28th
  International Conference on Financial Cryptography and Data Security}}.
  Springer.
\newblock


\bibitem[\protect\citeauthoryear{Miller, Xia, Croman, Shi, and Song}{Miller
  et~al\mbox{.}}{2016}]%
        {miller2016honey}
\bibfield{author}{\bibinfo{person}{Andrew Miller}, \bibinfo{person}{Yu Xia},
  \bibinfo{person}{Kyle Croman}, \bibinfo{person}{Elaine Shi}, {and}
  \bibinfo{person}{Dawn Song}.} \bibinfo{year}{2016}\natexlab{}.
\newblock \showarticletitle{The Honey Badger of BFT Protocols}. In
  \bibinfo{booktitle}{\emph{Proceedings of the 23rd ACM SIGSAC Conference on
  Computer and Communications Security}}. \bibinfo{pages}{31--42}.
\newblock


\bibitem[\protect\citeauthoryear{Most{\'e}faoui, Moumen, and
  Raynal}{Most{\'e}faoui et~al\mbox{.}}{2014}]%
        {mostefaoui2014signature}
\bibfield{author}{\bibinfo{person}{Achour Most{\'e}faoui},
  \bibinfo{person}{Hamouma Moumen}, {and} \bibinfo{person}{Michel Raynal}.}
  \bibinfo{year}{2014}\natexlab{}.
\newblock \showarticletitle{Signature-free Asynchronous Byzantine Consensus
  with $t<n/3$ and $O(n^2)$ Messages}. In \bibinfo{booktitle}{\emph{Proceedings
  of the 33rd ACM Symposium on Principles of Distributed Computing}}. ACM,
  \bibinfo{pages}{2--9}.
\newblock


\bibitem[\protect\citeauthoryear{Pease, Shostak, and Lamport}{Pease
  et~al\mbox{.}}{1980}]%
        {pease1980reaching}
\bibfield{author}{\bibinfo{person}{Marshall Pease}, \bibinfo{person}{Robert
  Shostak}, {and} \bibinfo{person}{Leslie Lamport}.}
  \bibinfo{year}{1980}\natexlab{}.
\newblock \showarticletitle{Reaching Agreement in the Presence of Faults}.
\newblock \bibinfo{journal}{\emph{J. ACM}} \bibinfo{volume}{27},
  \bibinfo{number}{2} (\bibinfo{year}{1980}), \bibinfo{pages}{228--234}.
\newblock


\bibitem[\protect\citeauthoryear{Peng, Zhang, Xu, Liu, Gao, Li, and Yu}{Peng
  et~al\mbox{.}}{2022}]%
        {peng2022neuchain}
\bibfield{author}{\bibinfo{person}{Zeshun Peng}, \bibinfo{person}{Yanfeng
  Zhang}, \bibinfo{person}{Qian Xu}, \bibinfo{person}{Haixu Liu},
  \bibinfo{person}{Yuxiao Gao}, \bibinfo{person}{Xiaohua Li}, {and}
  \bibinfo{person}{Ge Yu}.} \bibinfo{year}{2022}\natexlab{}.
\newblock \showarticletitle{Neuchain: A Fast Permissioned Blockchain System
  with Deterministic Ordering}.
\newblock \bibinfo{journal}{\emph{Proceedings of the VLDB Endowment}}
  \bibinfo{volume}{15}, \bibinfo{number}{11} (\bibinfo{year}{2022}),
  \bibinfo{pages}{2585--2598}.
\newblock


\bibitem[\protect\citeauthoryear{Qin, Wu, Amiri, Marcus, and Loo}{Qin
  et~al\mbox{.}}{2024}]%
        {qin2024bftgym}
\bibfield{author}{\bibinfo{person}{Haoyun Qin}, \bibinfo{person}{Chenyuan Wu},
  \bibinfo{person}{Mohammad~Javad Amiri}, \bibinfo{person}{Ryan Marcus}, {and}
  \bibinfo{person}{Boon~Thau Loo}.} \bibinfo{year}{2024}\natexlab{}.
\newblock \showarticletitle{BFTGym: An Interactive Playground for BFT
  Protocols}.
\newblock \bibinfo{journal}{\emph{Proceedings of the VLDB Endowment}}
  \bibinfo{volume}{17}, \bibinfo{number}{12} (\bibinfo{year}{2024}),
  \bibinfo{pages}{4261--4264}.
\newblock


\bibitem[\protect\citeauthoryear{Ramasamy and Cachin}{Ramasamy and
  Cachin}{2005}]%
        {ramasamy2005parsimonious}
\bibfield{author}{\bibinfo{person}{HariGovind~V Ramasamy} {and}
  \bibinfo{person}{Christian Cachin}.} \bibinfo{year}{2005}\natexlab{}.
\newblock \showarticletitle{Parsimonious Asynchronous Byzantine-fault-tolerant
  Atomic Broadcast}. In \bibinfo{booktitle}{\emph{Proceedings of the 9th
  International Conference On Principles of Distributed Systems}}. Springer,
  \bibinfo{pages}{88--102}.
\newblock


\bibitem[\protect\citeauthoryear{Reiter}{Reiter}{1994}]%
        {reiter1994secure}
\bibfield{author}{\bibinfo{person}{Michael~K Reiter}.}
  \bibinfo{year}{1994}\natexlab{}.
\newblock \showarticletitle{Secure Agreement Protocols: Reliable and Atomic
  Group Multicast in Rampart}. In \bibinfo{booktitle}{\emph{Proceedings of the
  2nd ACM Conference on Computer and Communications Security}}.
  \bibinfo{pages}{68--80}.
\newblock


\bibitem[\protect\citeauthoryear{Shrestha, Shrothrium, Kate, and
  Nayak}{Shrestha et~al\mbox{.}}{2025}]%
        {shrestha2024sailfish}
\bibfield{author}{\bibinfo{person}{Nibesh Shrestha}, \bibinfo{person}{Rohan
  Shrothrium}, \bibinfo{person}{Aniket Kate}, {and} \bibinfo{person}{Kartik
  Nayak}.} \bibinfo{year}{2025}\natexlab{}.
\newblock \showarticletitle{Sailfish: Towards Improving Latency of DAG-based
  BFT}. In \bibinfo{booktitle}{\emph{Proceedings of the 46th IEEE Symposium on
  Security and Privacy}}. IEEE.
\newblock


\bibitem[\protect\citeauthoryear{Spiegelman, Giridharan, Sonnino, and
  Kokoris-Kogias}{Spiegelman et~al\mbox{.}}{2022}]%
        {spiegelman2022bullshark}
\bibfield{author}{\bibinfo{person}{Alexander Spiegelman}, \bibinfo{person}{Neil
  Giridharan}, \bibinfo{person}{Alberto Sonnino}, {and}
  \bibinfo{person}{Lefteris Kokoris-Kogias}.} \bibinfo{year}{2022}\natexlab{}.
\newblock \showarticletitle{Bullshark: \text{DAG} \text{BFT} Protocols Made
  Practical}. In \bibinfo{booktitle}{\emph{Proceedings of the 29th ACM
  Conference on Computer and Communications Security}}. ACM,
  \bibinfo{pages}{2705--2718}.
\newblock


\bibitem[\protect\citeauthoryear{Wang, Duan, Clavin, and Zhang}{Wang
  et~al\mbox{.}}{2022}]%
        {wang2022bft}
\bibfield{author}{\bibinfo{person}{Xin Wang}, \bibinfo{person}{Sisi Duan},
  \bibinfo{person}{James Clavin}, {and} \bibinfo{person}{Haibin Zhang}.}
  \bibinfo{year}{2022}\natexlab{}.
\newblock \showarticletitle{BFT in Blockchains: From Protocols to Use Cases}.
\newblock \bibinfo{journal}{\emph{Comput. Surveys}} \bibinfo{volume}{54},
  \bibinfo{number}{10s} (\bibinfo{year}{2022}), \bibinfo{pages}{1--37}.
\newblock


\bibitem[\protect\citeauthoryear{Wu, Amiri, Asch, Nagda, Zhang, and Loo}{Wu
  et~al\mbox{.}}{2022}]%
        {wu2022flexchain}
\bibfield{author}{\bibinfo{person}{Chenyuan Wu},
  \bibinfo{person}{Mohammad~Javad Amiri}, \bibinfo{person}{Jared Asch},
  \bibinfo{person}{Heena Nagda}, \bibinfo{person}{Qizhen Zhang}, {and}
  \bibinfo{person}{Boon~Thau Loo}.} \bibinfo{year}{2022}\natexlab{}.
\newblock \showarticletitle{FlexChain: An Elastic Disaggregated Blockchain}.
\newblock \bibinfo{journal}{\emph{Proceedings of the VLDB Endowment}}
  \bibinfo{volume}{16}, \bibinfo{number}{1} (\bibinfo{year}{2022}),
  \bibinfo{pages}{23--36}.
\newblock


\bibitem[\protect\citeauthoryear{Wu, Amiri, Qin, Mehta, Marcus, and Loo}{Wu
  et~al\mbox{.}}{2024a}]%
        {wu2024towards}
\bibfield{author}{\bibinfo{person}{Chenyuan Wu},
  \bibinfo{person}{Mohammad~Javad Amiri}, \bibinfo{person}{Haoyun Qin},
  \bibinfo{person}{Bhavana Mehta}, \bibinfo{person}{Ryan Marcus}, {and}
  \bibinfo{person}{Boon~Thau Loo}.} \bibinfo{year}{2024}\natexlab{a}.
\newblock \showarticletitle{Towards Full Stack Adaptivity in Permissioned
  Blockchains}.
\newblock \bibinfo{journal}{\emph{Proceedings of the VLDB Endowment}}
  \bibinfo{volume}{17}, \bibinfo{number}{5} (\bibinfo{year}{2024}),
  \bibinfo{pages}{1073--1080}.
\newblock


\bibitem[\protect\citeauthoryear{Wu, Mehta, Amiri, Marcus, and Loo}{Wu
  et~al\mbox{.}}{2023}]%
        {wu2023adachain}
\bibfield{author}{\bibinfo{person}{Chenyuan Wu}, \bibinfo{person}{Bhavana
  Mehta}, \bibinfo{person}{Mohammad~Javad Amiri}, \bibinfo{person}{Ryan
  Marcus}, {and} \bibinfo{person}{Boon~Thau Loo}.}
  \bibinfo{year}{2023}\natexlab{}.
\newblock \showarticletitle{AdaChain: A Learned Adaptive Blockchain}.
\newblock \bibinfo{journal}{\emph{Proceedings of the VLDB Endowment}}
  \bibinfo{volume}{16}, \bibinfo{number}{8} (\bibinfo{year}{2023}),
  \bibinfo{pages}{2033--2046}.
\newblock


\bibitem[\protect\citeauthoryear{Wu, Zhu, and Hu}{Wu et~al\mbox{.}}{2024b}]%
        {wu2024blockchain}
\bibfield{author}{\bibinfo{person}{Honghu Wu}, \bibinfo{person}{Xiangrong Zhu},
  {and} \bibinfo{person}{Wei Hu}.} \bibinfo{year}{2024}\natexlab{b}.
\newblock \showarticletitle{A Blockchain System for Clustered Federated
  Learning with Peer-to-Peer Knowledge Transfer}.
\newblock \bibinfo{journal}{\emph{Proceedings of the VLDB Endowment}}
  \bibinfo{volume}{17}, \bibinfo{number}{5} (\bibinfo{year}{2024}),
  \bibinfo{pages}{966--979}.
\newblock


\bibitem[\protect\citeauthoryear{Xiao, Zhang, Lou, and Hou}{Xiao
  et~al\mbox{.}}{2020}]%
        {xiao2020survey}
\bibfield{author}{\bibinfo{person}{Yang Xiao}, \bibinfo{person}{Ning Zhang},
  \bibinfo{person}{Wenjing Lou}, {and} \bibinfo{person}{Y~Thomas Hou}.}
  \bibinfo{year}{2020}\natexlab{}.
\newblock \showarticletitle{A Survey of Distributed Consensus Protocols for
  Blockchain Networks}.
\newblock \bibinfo{journal}{\emph{IEEE Communications Surveys \& Tutorials}}
  \bibinfo{volume}{22}, \bibinfo{number}{2} (\bibinfo{year}{2020}),
  \bibinfo{pages}{1432--1465}.
\newblock


\bibitem[\protect\citeauthoryear{Yin, Malkhi, Reiter, Gueta, and Abraham}{Yin
  et~al\mbox{.}}{2019}]%
        {yin2019hotstuff}
\bibfield{author}{\bibinfo{person}{Maofan Yin}, \bibinfo{person}{Dahlia
  Malkhi}, \bibinfo{person}{Michael~K Reiter}, \bibinfo{person}{Guy~Golan
  Gueta}, {and} \bibinfo{person}{Ittai Abraham}.}
  \bibinfo{year}{2019}\natexlab{}.
\newblock \showarticletitle{\text{HotStuff}: \text{BFT} Consensus with
  Linearity and Responsiveness}. In \bibinfo{booktitle}{\emph{Proceedings of
  the 38th ACM Symposium on Principles of Distributed Computing}}.
  \bibinfo{pages}{347--356}.
\newblock


\bibitem[\protect\citeauthoryear{Zhang, Pan, Tijanic, and Jacobsen}{Zhang
  et~al\mbox{.}}{2024}]%
        {zhang2024prestigebft}
\bibfield{author}{\bibinfo{person}{Gengrui Zhang}, \bibinfo{person}{Fei Pan},
  \bibinfo{person}{Sofia Tijanic}, {and} \bibinfo{person}{Hans-Arno Jacobsen}.}
  \bibinfo{year}{2024}\natexlab{}.
\newblock \showarticletitle{Prestigebft: Revolutionizing View Changes in BFT
  Consensus Algorithms with Reputation Mechanisms}. In
  \bibinfo{booktitle}{\emph{2024 IEEE 40th International Conference on Data
  Engineering (ICDE)}}. IEEE, \bibinfo{pages}{1930--1943}.
\newblock


\bibitem[\protect\citeauthoryear{Zhang and Duan}{Zhang and Duan}{2022}]%
        {zhang2022pace}
\bibfield{author}{\bibinfo{person}{Haibin Zhang} {and} \bibinfo{person}{Sisi
  Duan}.} \bibinfo{year}{2022}\natexlab{}.
\newblock \showarticletitle{Pace: Fully Parallelizable BFT from Reproposable
  Byzantine Agreement}. In \bibinfo{booktitle}{\emph{Proceedings of the 29th
  ACM SIGSAC Conference on Computer and Communications Security}}.
  \bibinfo{pages}{3151--3164}.
\newblock


\end{thebibliography}

\appendix
\section{Correctness Analysis on $\Pi_{aaba}$}\label{sec:anal_aaba}
\subsection{Analysis on $\Pi_{aaba}$'s properties}
We prove that $\Pi_{aaba}$ correctly implements an AABA protocol.
\begin{revpara}
\subsubsection{Agreement.}
Nodes can only output at Line 20 or Line 26 in Algorithm~\ref{alg:aaba}.
If a correct node $p_i$ outputs $0$ at Line 20, i.e., outputs through the shortcut mechanism, it must have received $n-f$ \texttt{sho2} messages containing $0$.
Since $n \geq 3f+1$, any correct node must receive at least one \texttt{sho2} message containing $0$ and input $0$ to the subsequent ABA.
If another correct node $p_j$ outputs at Line 20, it will match $p_i$'s output.
Otherwise, based on ABA's validity and termination properties, $p_j$ will output $0$.
To sum up, as long as a correct node outputs $0$ through the shortcut mechanism, every correct node will eventually output $0$.

If $p_i$ outputs $b$ at Line 26, and $p_j$ has already output $0$ at Line 20, then, based on the above analysis, $p_i$ will also output $0$ (i.e., $b=0$).
Otherwise, by ABA's agreement property, $p_j$ will output $b$.

In conclusion, $\Pi_{aaba}$ achieves the agreement property.
\end{revpara}

\subsubsection{$1$-validity.}
If a correct node outputs $1$, it must output at Line 26 in Algorithm~\ref{alg:aaba}.
According to ABA's validity property, at least one correct node must input $1$ at Line 24.
Therefore, the set $S$ in Algorithm~\ref{alg:aaba} must include $1$, implying that at least one correct node, denoted as $p_i$, must have broadcast $\left<\texttt{sho1}, 1 \right>$.
Thus, $p_i$ must have received a message containing $\left<1, v, \sigma\right>$ such that $Q(v, \sigma)$=\texttt{true}.
In other words, some node must have inputted $\left<1, v, \sigma\right>$ s.t. $Q(v, \sigma)$=\texttt{true}.

\subsubsection{Biased-validity.}
If at least $f+1$ correct nodes input valid $\left<1, v, \sigma \right>$, each correct node will broadcast $\left<\texttt{sho1}, 1 \right>$ and will not broadcast $\left<\texttt{sho1}, 0 \right>$, according to Lines 3-10 in Algorithm~\ref{alg:aaba}.
Therefore, the set $S$ will only include $1$ and each correct node will input $1$ to the subsequent ABA protocol.
According to ABA's validity property, each correct node will output $1$.

\begin{revpara}
\subsubsection{Termination.}
\rev{If all correct nodes receive the input $0$, each correct node can broadcast $\left<\texttt{sho1}, 0 \right>$, provided it has not already broadcast $\left<\texttt{sho1}, 1 \right>$.
Conversely, if at least one correct node receives a valid $\left<1, v, \sigma \right>$, each correct node can broadcast $\left<\texttt{sho1}, 1 \right>$, provided it has not already broadcast $\left<\texttt{sho1}, 0 \right>$.}
In summary, each correct node can broadcast a \texttt{sho1} message.
Since $n-f \geq 2f+1$, there must be a bit, denoted as $b$, broadcast by at least $f+1$ correct nodes at either Line 6 or Line 10 of Algorithm~\ref{alg:aaba}.
According to Lines 11-13 of Algorithm~\ref{alg:aaba}, $b$ will be broadcast through the \texttt{sho1} message by all correct nodes.
Therefore, the set $S$ will eventually be non-empty, and it will definitely include $b$.

On the other hand, if a bit $c$ is included in a correct node's set $S$, at least $f+1$ correct nodes must have broadcast $\left<\texttt{sho1}, c \right>$, and eventually, every correct node will broadcast $\left<\texttt{sho1}, c \right>$. Therefore, each correct node will receive $\left<\texttt{sho1}, c \right>$ from at least $n-f$ nodes and include $c$ in its set $S$.

As a result, if a correct node broadcasts a message of $\left<\texttt{sho2}, b \right>$, this message will be accepted by every correct node, as $b$ is included in $S$.
Since $S$ will be non-empty, each correct node can broadcast a \texttt{sho2} message, which will be accepted by every correct node. 
Therefore, each correct node can receive at least $n-f$ valid \texttt{sho2} messages and subsequently input one bit into the ABA protocol.
According to ABA's termination property, each correct node will eventually output if it has not done so yet.
\end{revpara}

\subsubsection{Shortcut $0$-output.}
If all nodes input $0$, each correct node will only broadcast $\left<\texttt{sho1}, 0 \right>$ and receive only \texttt{sho1} messages containing $0$. Therefore, the set $S$ will only include $0$, and each correct node will only broadcast the \texttt{sho2} message containing $0$.
Thus, each correct node will receive $n-f$ valid \texttt{sho2} messages containing $0$ and output $0$ at Line 20.
This results in a shortcut latency of three communication rounds.

\subsection{Analysis on the early-stopping mechanism}\label{sec:anal-early-stop}
In this section, we mainly prove that $\Pi_{aaba}$ retains all AABA properties after adding the early-stopping mechanism. Besides, we prove that if $f+1$ correct nodes output $0$ at Line 20 in Algorithm~\ref{alg:aaba}, each correct node can output $0$ without finishing the subsequent ABA. 

First, since the early-stopping mechanism only works in the situation where $\Pi_{aaba}$ outputs $0$, it will not affect how $\Pi_{aaba}$ achieves the properties of $1$-validity and biased-validity. In other words, after adding the early-stopping mechanism, $\Pi_{aaba}$ can also achieve $1$-validity and biased-validity.
Besides, if a correct node outputs $0$ at Line 20 of Algorithm~\ref{alg:aaba}, any correct node that outputs from Algorithm~\ref{alg:early-stop} can only output $0$, which maintains the agreement property.
If no correct node exits from Algorithm~\ref{alg:early-stop}, all correct nodes will participate in the ABA protocol and eventually terminate after outputting from ABA.
On the contrary, if a correct node exits from Algorithm~\ref{alg:early-stop}, it must have received $n-f$ \texttt{stop} messages, at least $n-2f$ ($n-2f \geq f+1$) of which are broadcast from correct nodes. Therefore, each correct node can receive at least $f+1$ \texttt{stop} messages and will also broadcast a \texttt{stop} message if it has not broadcast one.
Eventually, each correct node can receive $n-f$ \texttt{stop} messages and exit from Algorithm~\ref{alg:early-stop}. Thus, $\Pi_{aaba}$ can still achieve the termination property after adding the early-stopping mechanism.
Furthermore, if all nodes input $0$, each correct node can output $0$ at Line 20, leading to a latency of three communication rounds, regardless of whether the early-stopping mechanism is added or not.
To sum up, $\Pi_{aaba}$ can achieve all properties of AABA after adding the early-stopping mechanism.

If $f+1$ correct nodes output $0$ at Line 20 in Algorithm~\ref{alg:aaba}, each of these nodes will broadcast a \texttt{stop} message in Algorithm~\ref{alg:early-stop}.
Then, each correct node can receive $f+1$ \texttt{stop} messages and broadcast a \texttt{stop} message if it has not broadcast one.
Therefore, each correct node can receive $n-f$ \texttt{stop} messages and then exit from the $\Pi_{aaba}$ instance without finishing the subsequent ABA protocol.

\section{Other Related Works}\label{sec:other-relat-work}
The synchronous network assumption, on the one hand, simplifies the design of protocols and, on the other hand, enhances fault tolerance. As a result, many early works~\cite{pease1980reaching, reiter1994secure, kihlstrom1998securering}, as well as some recent ones~\cite{chan2018pili, abraham2020sync}, are based on this assumption. 
Synchronous networks typically assume that messages between correct nodes are guaranteed to be delivered within a certain period, denoted by $\Delta$. However, setting the appropriate value for $\Delta$ presents a significant challenge. If $\Delta$ is set too small, the assumption is likely to be violated, compromising the protocol’s safety. Conversely, if $\Delta$ is set too large, nodes will have to wait longer before proceeding to the next step, thereby reducing the protocol's performance.

To avoid the safety issues present in the synchronous network assumption, Dwork et al.~\cite{dwork1988consensus} proposed the partially-synchronous network assumption. Under this assumption, the system transitions into a synchronous state after an unknown duration during which the network may be asynchronous. BFT protocols designed based on the partially-synchronous assumption do not rely on the assumption's correctness for safety but only for liveness. Since the introduction of the partially-synchronous assumption, it has been adopted by numerous protocols~\cite{gai2023scaling, zhang2024prestigebft, kang2024spotless}, with PBFT being the most representative~\cite{castro1999practical}.
Many subsequent works have optimized the PBFT protocol, such as introducing a fast path~\cite{kotla2007zyzzyva, gueta2019sbft, dai2022trebiz}, assuming an optimistic case~\cite{buchnik2020fireledger}, utilizing trusted hardware~\cite{liu2018scalable, arun2022scalable}. The rise of blockchain technology~\cite{wu2023adachain, wu2024blockchain, wu2024towards} also inspired researchers to integrate block and chain data structures into consensus protocol design, leading to notable partially-synchronous protocols like Tendermint~\cite{amoussou2019dissecting} and HotStuff~\cite{yin2019hotstuff}.
Qin et al. designed an interactive platform to evaluate the partially-synchronous BFT protocols~\cite{qin2024bftgym}, but it is not suitable for our work.

Despite their long-standing popularity, partially-synchronous protocols were shown by Miller et al.~\cite{miller2016honey} to be vulnerable to network attacks that can result in a loss of liveness. As a result, recent researchers, including our work, have shifted the focus to asynchronous networks.
Furthermore, partially-synchronous protocols typically introduce a timer, which is hard to tune.


\rev{}
\end{document}